\documentclass[a4paper,12pt]{iopart}
\usepackage{iopams}

\usepackage[colorlinks=true,linkcolor=blue,citecolor=blue,urlcolor=blue]{hyperref}
\usepackage{bbm}
\usepackage{multirow}

\usepackage{tocloft}

\newcommand{\eqref}{\eref}
\newcommand{\eqr}{\eref}

\usepackage{graphicx,psfrag}

\newcommand{\dd}{{\rmd}}
\newcommand{\ee}{{\rme}}
\newcommand{\ii}{{\rmi}}
\newcommand{\sign}{{\rm sign}}

\newcommand{\nbs}{n_{\rm BS}}

\newcommand{\chor}{{c_{\rm H}}}

\newcommand{\kb}{\kappa_{\rm B}}
\newcommand{\kw}{\kappa_{\rm W}}

\newcommand{\bx}{{\bf x}}

\newcommand{\nx}{\nabla_{\bf x}}
\newcommand{\pdx}{\partial_x}
\newcommand{\pdt}{\partial_t}
\newcommand{\re}{{\rm Re}}
\newcommand{\im}{{\rm Im}}
\newcommand{\tra}{{\rm T}}
\newcommand{\dx}{\dd x}
\newcommand{\dom}{\dd\omega}
\newcommand{\hbm}{\frac{\hbar^2}{2m}}

\newcommand{\om}{\omega}
\newcommand{\ommax}{\omega_{\rm max}}
\newcommand{\omp}{\omega'}
\newcommand{\la}{\lambda_a}

\newcommand{\las}{\lambda_a^*}

\newcommand{\hP}{\hat\Psi}
\newcommand{\Pn}{\Psi_0}
\newcommand{\rn}{\rho_0}
\newcommand{\rnx}{\rho_0(x)}
\newcommand{\rnxp}{\rho_0(x')}
\newcommand{\hr}{\hat\rho_1}
\newcommand{\hH}{\hat H}
\newcommand{\hHe}{\hat H_{\rm eff}}
\newcommand{\hHr}{\hat H_{\rm r}}
\newcommand{\hHc}{\hat H_{\rm c}}
\newcommand{\hHm}{\hat H_{\rm mix}}
\newcommand{\hp}{\hat\phi}
\newcommand{\hPd}{\hat\Psi^\dagger}
\newcommand{\hpd}{\hat\phi^\dagger}
\newcommand{\hc}{\hat c}
\newcommand{\hpds}{\hat\phi^{\dagger 2}}
\newcommand{\hW}{\hat W}
\newcommand{\ha}{\hat a}
\newcommand{\hb}{\hat b}
\newcommand{\hch}{\hat \chi}
\newcommand{\had}{\hat a^\dagger}

\newcommand{\wom}{ W_\omega^\alpha}
\newcommand{\womb}{ \bar W_\omega^\alpha}
\newcommand{\aom}{\hat a_\omega^\alpha}
\newcommand{\aomd}{\hat a_\omega^{\alpha\dagger}}

\newcommand{\vl}{V_{a}}
\newcommand{\vlb}{\bar V_{a}}
\newcommand{\bl}{\hat b_{a}}
\newcommand{\bld}{\hat b_{a}^\dagger}

\newcommand{\zl}{Z_{a}}
\newcommand{\zlb}{\bar Z_{a}}
\newcommand{\cl}{\hat c_{a}}
\newcommand{\cld}{\hat c_{a}^\dagger}

\newcommand{\wlpl}{W_{a+}}
\newcommand{\wlplb}{\bar W_{a+}}
\newcommand{\dlpl}{\hat d_{a+}}
\newcommand{\dlpld}{\hat d_{a+}^\dagger}

\newcommand{\wlmi}{W_{a-}}
\newcommand{\wlmib}{\bar W_{a-}}
\newcommand{\dlmi}{\hat d_{a-}}
\newcommand{\dlmid}{\hat d_{a-}^\dagger}

\newcommand{\womp}{ W_{\omega'}^{\alpha'}}
\newcommand{\wombp}{ \bar W_{\omega'}^{\alpha'}}
\newcommand{\aomp}{\hat a_{\omega'}^{\alpha'}}
\newcommand{\aomdp}{\hat a_{\omega'}^{\alpha'\dagger}}

\newcommand{\vlp}{V_{a'}}
\newcommand{\vlbp}{\bar V_{{a'}}}

\newcommand{\bldp}{\hat b_{{a'}}^\dagger}

\newcommand{\zlp}{Z_{{a'}}}
\newcommand{\zlbp}{\bar Z_{{a'}}}

\newcommand{\cldp}{\hat c_{{a'}}^\dagger}

\newcommand{\wlplp}{W_{{a'}+}}
\newcommand{\wlplbp}{\bar W_{{a'}+}}

\newcommand{\dlpldp}{\hat d_{{a'+}}^\dagger}

\newcommand{\wlmip}{W_{{a'}-}}
\newcommand{\wlmibp}{\bar W_{{a'}-}}

\newcommand{\dlmidp}{\hat d_{{a'}-}^\dagger}

\newcommand{\pom}{\phi_\omega^\alpha(x)}
\newcommand{\poms}{(\phi_\omega^{\alpha}(x))^*}
\newcommand{\vpom}{\varphi_\omega^\alpha(x)}
\newcommand{\vpoms}{(\varphi_\omega^{\alpha}(x))^*}
\newcommand{\pomp}{\phi_{\omega'}^{\alpha'}(x)}

\newcommand{\vpomp}{\varphi_{\omega'}^{\alpha'}(x)}
\newcommand{\vpomsp}{(\varphi_{\omega'}^{\alpha'}(x))^*}

\newcommand{\xl}{\xi_{a}(x)}
\newcommand{\xls}{(\xi_{a}(x))^*}
\newcommand{\el}{\eta_{a}(x)}

\newcommand{\elsx}{\eta_{a}}

\newcommand{\xlsx}{\xi_{a}}

\newcommand{\els}{(\eta_{a}(x))^*}
\newcommand{\psl}{\psi_{a}(x)}

\newcommand{\zel}{\zeta_{a}(x)}
\newcommand{\zels}{(\zeta_{a}(x))^*}

\newcommand{\plplsx}{\phi_{a+}}

\newcommand{\vplplsx}{\varphi_{a+}}

\newcommand{\plpl}{\phi_{a+}(t,x)}

\newcommand{\vplpl}{\varphi_{a+}(t,x)}
\newcommand{\vplpls}{(\varphi_{a+}(t,x))^*}

\newcommand{\plmisx}{\phi_{a-}}

\newcommand{\vplmisx}{\varphi_{a-}}

\newcommand{\plmi}{\phi_{a-}(t,x)}

\newcommand{\vplmi}{\varphi_{a-}(t,x)}
\newcommand{\vplmis}{(\varphi_{a-}(t,x))^*}

\newcommand{\pslp}{\psi_{{a'}}(x)}

\newcommand{\zelp}{\zeta_{{a'}}(x)}

\newcommand{\com}{\chi_\omega^\alpha(x)}

\newcommand{\clpl}{\chi_{a+}(t,x)}

\newcommand{\clmi}{\chi_{a-}(t,x)}

\newcommand{\sil}{\sigma_a(x)}

\newcommand{\nl}{\nu_a(x)}

\newcommand{\comsxp}{(\chi_\omega^{\alpha}(x'))^*}

\newcommand{\clplsxp}{(\chi_{a+}(t',x'))^*}

\newcommand{\clmisxp}{(\chi_{a-}(t',x'))^*}

\newcommand{\silsxp}{(\sigma_a(x'))^*}

\newcommand{\nlsxp}{(\nu_a(x'))^*}

\newcommand{\kone}{k_\om^{(1)}}
\newcommand{\ktwo}{k_\om^{(2)}}

\newcommand{\ppl}{\ee^{\ii\theta_+}}
\newcommand{\pmi}{\ee^{\ii\theta_-}}
\newcommand{\pplm}{\ee^{-\ii\theta_+}}
\newcommand{\pmim}{\ee^{-\ii\theta_-}}

\newcommand{\php}[1]{\ee^{+\ii #1}}
\newcommand{\phm}[1]{\ee^{-\ii #1}}

\newcommand{\spin}[2]{\pmatrix{ {#1} \cr {#2}}}
\newcommand{\scal}[2]{{\langle #1 | #2 \rangle}}
\newcommand{\comm}[2]{[#1,#2]}
\newcommand{\vev}[1]{{\langle 0 | #1 | 0 \rangle}}

\newcommand{\urlnjp}{\href{http://stacks.iop.org/NJP/12/095015/mmedia}{stacks.iop.org/NJP/12/095015/mmedia}}

\renewcommand{\mailto}[1]{{\href{mailto:#1}{\tt #1}}}

\begin{document}

\title{Black hole lasers in Bose--Einstein condensates}

\author{S Finazzi$^1$ and R Parentani$^2$}
\address{
$^1$~SISSA, via Bonomea 265, 34136 Trieste, Italy and INFN, sezione di Trieste, Italy\\
$^2$~Laboratoire de Physique Th\'eorique, CNRS UMR 8627, B\^at. 210, Universit\'e Paris-Sud 11, 91405 Orsay Cedex, France
}

\eads{\mailto{finazzi@sissa.it} and \mailto{renaud.parentani@th.u-psud.fr}}

\begin{abstract}
We consider elongated condensates that cross twice the speed of sound.
In the absence of periodic boundary conditions, the phonon spectrum possesses a discrete and finite set of complex frequency modes that induce a laser effect. This effect constitutes a dynamical instability and is due to the fact that the supersonic region acts as a resonant cavity.
We numerically compute the complex frequencies and density--density correlation function. We obtain patterns with very specific signatures.
In terms of the gravitational analogy, the flows we consider correspond to a pair of black hole and white hole horizons, and the laser effect can be conceived as a self-amplified Hawking radiation. This is verified by comparing the outgoing flux at early time with the standard black hole radiation.
\\

\noindent Online supplementary data available from~\urlnjp
\end{abstract}

\submitto{\NJP}

\noindent Accepted: 11 Jun 2010.\\
\noindent \url{http://stacks.iop.org/1367-2630/12/095015} 

\newpage

\setcounter{tocdepth}{2}
\section*{Contents}
\vspace{-30pt}
\tableofcontents

\section{Introduction}

In 1981, Unruh suggested~\cite{unruh} that the analogue of black hole radiation could be observed in a quantum fluid that crosses the speed of sound, thereby forming a sonic horizon. Since then, several types of fluids and several types of flows were considered~\cite{lr}.
Very recently, a near-stationary supersonic flow engendering a pair of horizons (a black hole one followed by a white hole one) was realized in a Bose--Einstein condensate~\cite{tech}.
In such a background, because of the anomalous dispersion of Bogoliubov phonons, one expects to get a kind of laser effect~\cite{cj}--\cite{cp} due to a self-amplification of the Hawking radiation.

The first aim of this paper is to provide the theoretical basis of this effect starting from the Bogoliubov--de Gennes equation, i.e. without making use of the gravitational analogy.
In this we apply to flows containing two horizons the treatment of~\cite{mp} that was only applied to a single (black hole or white hole) sonic horizon.
Following~\cite{cp} we then establish that the laser effect is governed by a discrete set of complex frequency modes that correspond to the resonant modes of the cavity formed by the region bordered by the two horizons.
In a classical description, this discrete set governs the dynamical instability of the flow.
The second aim is to compute the spectrum of the complex frequency modes
by theoretical and numerical methods.
The excellent agreement of the results validates both the concepts and the semi-classical methods used in the theoretical approach.
Finally we consider two observables: the mean number of emitted phonons, and the two-point function of the phonon field
that governs the density--density correlation pattern~\cite{carusotto1,carusotto2}. At sufficiently late time,
both observables are governed by a single mode, the most unstable one.
In this late time regime, their behaviours are identical to those one would obtain using a classical description
of the density perturbations.
At early time instead, when starting from vacuum configurations, correspondence with the quantum phonon flux emitted by a black hole horizon~\cite{mp} is established.

The propagation of phonons in flows containing two horizons has already been considered. However, toroidal configurations with periodic boundary conditions were generally used~\cite{Garay,Gardiner}.
In that case, the spectrum is very complicated because it results from a combination of two effects: the discreteness of the wave vectors defined on the torus interferes with that associated with modes that are trapped in the supersonic region.
As a result, not only the analysis is difficult, but the relationships with the Hawking effect and the black hole laser effect~\cite{cj}--\cite{cp} are hard to draw.
On the contrary, when dealing with continuous wave vectors, the analysis of the complex frequency mode is simpler, and the relationship with Hawking radiation is easily made.

We have organized this paper as follows.
In~\sref{sec:framework} we present the Bogoliubov--de Gennes equation in a way that is suitable for our aims.
In~\sref{sec:theory} we determine the spectrum of asymptotically bound modes, and explain why complex frequency modes appear when the speed of sound is twice crossed.
We then compute the eigen--frequencies using semi-classical methods.
In~\sref{sec:numerics} we numerically solve the Bogoliubov--de Gennes equation and compute the spectrum.
We then evaluate the mean occupation number and compare it with the spectrum
one would obtain if only the black hole horizon were present.
We finally evaluate the correlation pattern of the density--density two-point function.

\section{Framework}
\label{sec:framework}

\subsection{Black-hole--white-hole geometries} 
\label{sec:bhlasers}

Using the analogy~\cite{unruh,lr} between the wave equation governing sound propagation in a moving fluid and that governing light propagation in a curved space-time, one can associate an acoustic geometry to the fluid flow.
In 1+1 dimensions, these geometries are of the form~\cite{cj}
\begin{equation}\label{eq:metric}
 \dd s^2=-c^2 \dd t^2+(\dd x-v\dd t)^2,
\end{equation}
where $v$ is the flow velocity and $c$ the speed of sound.
In this language, a black-hole--white-hole geometry is obtained when $v$ crosses twice $c$. Indeed, a sonic horizon is present whenever $v$ crosses $c$.
Assuming that the fluid flows from right to left ($v<0$), the location $x_{\rm H}$ of a horizon is given by $c(x_{\rm H})+v(x_{\rm H})=0$. In our case, we set the location of the white hole horizon and that of the black hole respectively to $x_{\rm W}=-L$ and $x_{\rm B}=L$.
These divide the $x$-axis in three regions, which we call I on the left of the white horizon ($x<-L$), II between the two horizons ($-L<x<L$) and III on the right of the black horizon ($x>L$). The flow is supersonic in the internal region II and subsonic otherwise. The opposite case where the velocity is subsonic inside does not give rise to a laser effect, and will be considered elsewhere.

\begin{figure}
 \psfrag{q}[c][c]{\small $q$}
 \psfrag{d}[c][c]{\small $D$}
 \psfrag{x}[c][c]{\small $\kappa x/\chor$}
 \psfrag{vc}[c][c]{\small $|v|/\chor,\,c/\chor)$}
 \begin{flushright}
  \includegraphics[width=0.85\textwidth]{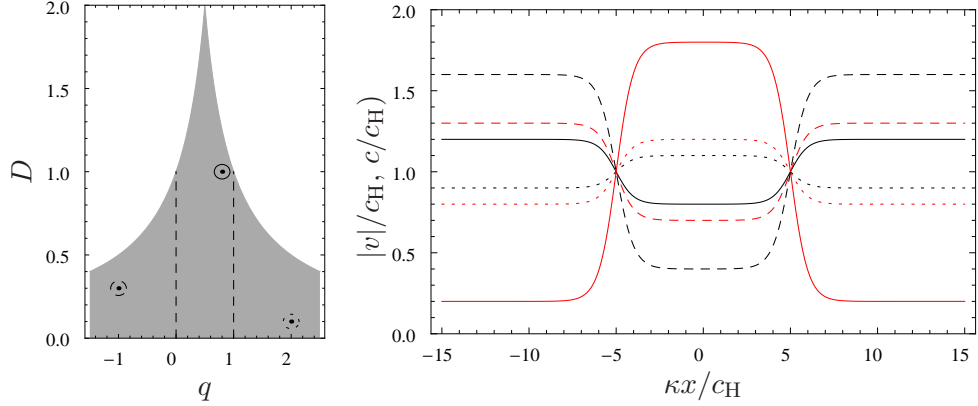}
 \end{flushright}
 \caption{Left panel: allowed values for $q$ and $D$ giving rise to a black-hole--white-hole geometry.
Right panel: Velocity profiles $c(x)/\chor$ (black lines) and $|v(x)|/\chor$ (red lines) for three different values of $(q,D)$: $q=0.8$, $D=1$ (solid lines), $q=-1$, $D=0.3$ (dashed lines) and $q=2$, $D=0.1$ (dotted lines). These values are also reported in the left plot. All profiles with $\kw=\kb=\kappa$.}
 \label{fig:qD}
\end{figure}

In the present work, our analysis is restricted to the following stationary profiles:
\begin{equation}\label{eq:velocity}
 \fl
 c(x)+v(x)=\chor D\, \sign(x^2-L^2) 
 \tanh^{1/n}\left[\left(\frac{\kw|x+L|}{\chor D}\right)^n\right] \tanh^{1/n}\left[\left(\frac{\kb|x-L|}{\chor D}\right)^n\right],
\end{equation}
which generalize to two horizons those used in~\cite{mp}. $\chor$ is the sound speed at the horizons, $2L$ is the distance between them, $D$ determines the size of the near-horizon regions where the metric is not flat, $n$ controls the sharpness of the transition to the flat regions, and $\kw$ and $\kb$ control the surface gravities.
Indeed, the surface gravity is defined by~\cite{lr}
\begin{equation}
 g_{\rm s}=\frac{1}{2}\left.\frac{\dd(c^2-v^2)}{\dd x}\right|_{x=x_{\rm H}},
\end{equation}
which yields $-\chor\kw$ for the white hole and $\chor\kb$ for the black hole.
The metric~\eref{eq:metric} is then completely fixed by introducing an extra parameter $q$:
\begin{eqnarray}\label{eq:cv}
\eqalign{
 c(x)=\chor+(1-q)[c(x)+v(x)], \\
 v(x)=-\chor+q[c(x)+v(x)],
 }
\end{eqnarray}
which specifies how $c+v$ is shared between $c$ and $v$.
When restricting to the cases where
\begin{equation}
 c(x)>0,\quad v(x)<0,\quad D>0,
\end{equation}
the range of $D$ and $q$ is limited.
The allowed values are graphically illustrated by the shaded area in~\fref{fig:qD}, left panel.
Three velocity profiles corresponding to different couples $(q,D)$ are plotted in~\fref{fig:qD}, right panel.
%

\subsection{Density perturbations in Bose Einstein condensates} 
\label{sec:BEC}

We present the main steps leading to the equations for linear density perturbations in Bose--Einstein condensates~\cite{Dalfovo}, having in mind cases where the condensate crosses the speed of sound. We follow~\cite{mp}, where more details can be found.

At low temperature, the quantum properties of a gas of weakly interacting atoms are efficiently described by a second quantized field satisfying the commutation relation
\begin{equation}\label{eq:comm}
 \comm{\hP(t,\bx)}{\hPd(t,\bx')}=\delta^3(\bx-\bx'),
\end{equation}
and by its Hamiltonian
\begin{equation}\label{eq:Hsc}
 \hH = \int\! \dd^3{x} \left\{\frac{\hbar^2}{2m} \nx \hPd \, \nx\hP + V  \hPd\hP  + \frac{g}{2} 
 \hPd\hPd\hP\hP    \right\},
\end{equation}
where $m$ is the atom mass, $V$ the external potential and $g$ the effective coupling.
The last two quantities can depend on both $t$ and $\bx$.
When a significant fraction of the atoms condense, it is meaningful to expand $\hP$ in a $c$-number function $\Pn$, describing the condensed part, plus a field operator $\hp$, describing (relative) density perturbations over the condensate
\begin{equation}\label{eq:defphi}
 \hP=\Pn(1+\hp).
\end{equation}

In what follows, we consider elongated condensates, which means that the transverse excitations have sufficiently high energies that they are not excited. In this case, $\Pn$ and $\hp$ are effectively one-dimensional fields.
For simplicity, we also assume that the condensate is infinitely long, in order to avoid discussing the discreteness of the longitudinal wave number $k$.
The discrete case is indeed more complicated, and treated in details in~\cite{Gardiner}.
Finally, we assume that the condensate is stationary, which means that $\Pn$ is of the form
\begin{equation}
 \Pn(t,x)=\ee^{-\ii\mu t/\hbar}\times\sqrt{\rn(x)}\ee^{\ii\theta_0(x)},
\end{equation}
where $\mu$ is the chemical potential, $\rn(x)$ the mean density of condensed atoms and
\begin{equation}
 v(x)=\frac{\hbar}{m}\pdx\theta_0(x)
\end{equation}
is their mean velocity.
In this case, the Gross--Pitaevskii equation
\begin{equation}
 \ii\hbar\pdt\Pn=\left[-\hbm\pdx^2+V+g\rn\right]\Pn
\label{eq:GP}
\end{equation}
reduces to
\begin{equation}
 \mu = \frac{1}{2}mv^2-\hbm\frac{\pdx^2\sqrt{\rn}}{\rn}+V+g\rn,\qquad
 \pdx(v\rn)=0,\label{eq:cont}
\end{equation}
where the second equation is the continuity equation for a stationary flow.

At the linear level, $\hp$ obeys the Bogoliubov-de Gennes equation. In the present case, using \eqr{eq:defphi} and \eqr{eq:cont}, this equation reads
\begin{equation}
 \ii\hbar\pdt\hp = \left[T_\rho-\ii v\hbar\pdx + m c^2 
\right]
\hp + m c^2 
\hpd,\label{eq:hp}
\end{equation}
where we introduced the $x$-dependent speed of sound
\begin{equation}
 c^2(x)\equiv\frac{g(x)\rn(x)}{m},
\end{equation}
and where
\begin{equation}\label{eq:Trho}
 T_\rho\equiv -\frac{\hbar^2}{2m} \frac {1}{\rn}\pdx\rn\pdx=-\frac{\hbar^2 }{2m} \, v\pdx \frac{1}{v}\pdx
\end{equation}
reduces to the usual kinetic operator when the background condensate is homogeneous.
The second expression is only valid in stationary flows.
Equation \eqr{eq:hp} tells us that the perturbation $\hp$ is coupled to the condensate only through the functions $v$ and $c$.
The disappearance, from \eqr{eq:hp}, of the potential $V$, the quantum potential $-\hbm\frac{\pdx^2\sqrt{\rn}}{\rn}$, and the coupling $g$ constitutes an essential step towards the notion of the acoustic metric of \eqr{eq:metric}; see~\cite{mp} for more details.

Using the canonical commutation relation
\begin{equation}\label{eq:commphi}
 \comm{\hp(t,x)}{\hpd(t,x')}=\frac{1}{\rn(x)}\delta(x-x'),
\end{equation}
which follows from equations~\eref{eq:comm} and~\eref{eq:defphi}, \eref{eq:hp} is the Heisenberg equation
\begin{equation}\label{eq:phi}
 \ii\hbar\pdt\hp = \comm{\hp}{\hHe},
\end{equation}
engendered by the Hermitian Hamiltonian
\begin{equation}\label{eq:heffsym}
\fl 
\hHe=
\int\! \dx\,\frac{\rn}{2}
\left\{\hpd\left[T_\rho-\ii v\hbar\pdx\right]\hp +[(T_\rho+\ii v\hbar\pdx)\hpd]\hp 
+g{\rn}\left(\hpds+\hp^2+ 2 \hpd \hp \right)\right\},
\end{equation}
which is obtained by expanding \eref{eq:Hsc} to second order in $\hp$, $\hpd$;
see~appendix~\hyperlink{app:motion}{A}.

\section{Theoretical analysis}
\label{sec:theory}

\subsection{Spectrum of bound modes and quantization}
\label{sec:quantization}

In stationary condensates, the spectrum of \eqr{eq:heffsym} can be characterized using very general properties.
These are first, the fact that the eigenmodes are asymptotically bound (to guarantee that they have a well-defined norm
since the spatial domain of $\hp$ is infinite), second, the Hermiticity of \eqr{eq:heffsym}, i.e. its self-adjointness with respect to the scalar product defining the eigen-modes norm, and third, that this scalar product is {\it not} positive definite, as it is the case here, see~\eqr{eq:scalar}, and for bosonic fields, see e.g.~\cite{Greinerbook}.

When these conditions are met, the spectrum of \eqr{eq:heffsym} contains a continuous set of real frequency modes labelled by $\om$ and a discrete index $\alpha$, plus, {\it possibly}, a discrete and finite set of complex frequency modes that appear in pairs with complex conjugated frequencies.
(If the scalar product were positive definite, as is the case for fermions,
the spectrum of \eqr{eq:heffsym} would be purely real.)
Throughout this paper, complex frequencies shall be written as $\la =\om_a + i \Gamma_a $, where $a$ is a positive integer which labels the discrete set of pairs, and where $\om_a$ and $\Gamma_a $ are both real and positive.
The other cases are reached by complex conjugation and/or multiplication by $-1$.
The discrete index $\alpha$ describes the subset of modes with the same real frequency.
As explained below, in one dimension, it contains two or four modes depending on whether the flow is subsonic or supersonic.

As a result, when propagating in a stationary condensate, $\hp$, obeying \eqr{eq:commphi} and \eqr{eq:phi}, can always be expanded (see~appendix~\hyperlink{app:quantization}{B}) as
\begin{eqnarray}\label{eq:phiexpansion}
 \fl \quad 
\hp(t,x) =\int\!\dom\sum_{\alpha}\left[\phm{\om t}\pom\aom+\php{\om t}\vpoms\aomd\right]
\nonumber \\ \fl
 \quad \quad \quad +\sum_a\left[\phm{\la t}\xl\bl+\phm{\las t}\psl\cl+\php{\las t}\els\bld+\php{\la t}\zels\cld\right],
\end{eqnarray}
where the modes satisfy the normalization relations
\begin{eqnarray}
 \int\!\dx\,\rn\left[\poms\pomp-\vpoms\vpomp\right]=\delta_{\alpha\alpha'}\delta(\om-\omp),\label{eq:normphi}\\
 \int\!\dx\,\rn\left[\xls\pslp-\els\zelp\right]=\ii\delta_{\lambda_a\lambda_{a'}}.\label{eq:normpsi}
\end{eqnarray}
Correspondingly, the operators $\ha$, $\hb$, $\hc$ satisfy the commutation relations
\begin{eqnarray}
 \comm{\aom}{\aomdp}=\delta_{\alpha\alpha'}\delta(\om-\omp),\\
 \comm{\bl}{\cldp}=\ii\delta_{\lambda_a\lambda_{a'}}.
 \label{eq:combcd}
\end{eqnarray}
All the other scalar products, and all other commutators, vanish. In particular, one has $\comm{\bl}{\bldp}= \comm{\cl}{\cldp}= 0$.
Hence $\hb_a$ and $\hc_a$ are not destruction operators. In fact, the operators $\hb_a$ and $\hc_a$ form pairs~\cite{cp} that characterize complex, i.e. with two degrees of freedom, harmonic oscillators that are unstable, and for which therefore there is no notion of quanta. (Concomitantly, one verifies that the norm $\int\!\dx\,\rn\left[\vert \xlsx \vert^2- \vert \elsx \vert^2\right]$ vanishes.)
It should be noticed that the eigenfrequencies associated with $\hb_a$ and $\hc_a$ are complex conjugated from each other (respectively $\la$ and $\la^*$), something guaranteed by the Hermiticity of~\eqr{eq:heffsym}.

In terms of the above eigenmodes and operators, \eqr{eq:heffsym} reads
\begin{eqnarray}\label{eq:hamiltonianabc}
 \hHe = \int\!\dom\sum_\alpha\hbar\om\left(\aomd\aom-\int\!\dx\,\rnx|\vpom|^2  \right)
 \nonumber\\
 \quad  \quad \quad +\sum_a \ii \left[\hbar\las\left(\bld\cl+\int\!\dx\,\rnx\zel\els\right)-
\mbox{h.c.} \right].
\end{eqnarray}
The first line contains the usual sum over harmonic oscillators of real frequency and the $c$-number term accounting for the depletion, while the second line is due to the complex frequency modes.
It should be noticed that the unusual form of the second $c$-number term follows from the unusual scalar product of \eqr{eq:normpsi}.
This Hamiltonian is not bounded from below due to the $b_a^\dagger c_a$ terms.
Hence the vacuum can no longer be defined as the ground state of $H$,
as one might have expected since one is dealing with unstable oscillators.
This gives rise to some ambiguity when choosing the initial `vacuum' state (see~\sref{sec:correlations}).

Eqs. (\ref{eq:phiexpansion}) and (\ref{eq:hamiltonianabc}) can be rewritten in a more familiar form by decomposing each couple ($\hb_a, \hc_a$) into two couples of destruction/creation operators $\dlpl$, $\dlpld$ and $\dlmi$, $\dlmid$, as shown in appendix~\hyperlink{app:quantization}{B}. The field becomes
\begin{eqnarray}
\fl  \hp(t,x)=\int\!\dom\sum_{\alpha}\left[\phm{\om t}\pom\aom+\php{\om t}\vpoms\aomd\right]
 \nonumber\\  \quad \quad
 \fl+\sum_a\left[\plpl\,  \dlpl+\plmi\, \dlmi+\vplpls\, \dlpld+\vplmis\, \dlmid\right], 
\label{eq:phiadd}
\end{eqnarray}
and the Hamiltonian~\eqref{eq:hamiltonianabc}
\begin{eqnarray}
 \fl\hHe &=& \int\!\dom\sum_\alpha\hbar\om\left(\aomd\aom-\int\!\dx\,\rnx|\vpom|^2  \right)
  \nonumber\\
 \fl&&+\sum_a\hbar\om_a\left[\dlpld\dlpl-\dlmid\dlmi-\int\!\dx\,\rnx\left(|\vplpl|^2-|\vplmi|^2\right)\right]
 \nonumber\\
 \fl&&+
\sum_a\ii\hbar\Gamma_a\left[
\dlpld\dlmid 
  +\int\!\dx\,\rnx\left(
\plpl\vplmi 
\right) - \mbox{h.c.} \right].
\label{eq:hamiltonianadd}
\end{eqnarray}

We conclude with two remarks.
Firstly, if one can describe the discrete set using the $\hat d_a$ operators, there is a price to pay as the associated modes $\plplsx$, $\vplplsx$, $\plmisx$, $\vplmisx$ are not frequency eigenmodes. Hence their time dependence cannot be factorized.
Secondly, in usual circumstances, the discrete set of complex frequency modes is empty, as is the case in homogeneous and in near-homogeneous condensates. Nevertheless, when the condensate crosses twice the speed of sound, the discrete set is (generally) nonempty.
To explain why, we first need to study the spectrum in homogeneous subsonic and supersonic flows, and then understand how to paste the corresponding modes when the flow crosses the speed of sound.
When solving this, the dispersive properties of $\hp$ become essential, as we now recall, and as first understood in~\cite{cj}.

\subsection{Mode analysis}
\label{sec:homog}

Inserting the field expansion~\eqref{eq:phiexpansion} in~\eqref{eq:hp}, one obtains a system of two $c$-number equations
\begin{equation}
\eqalign{
 \left[\hbar(\lambda +\ii v\pdx) - T_\rho - mc^2\right]\phi_\lambda = mc^2\varphi_\lambda,\\
 \left[-\hbar(\lambda +\ii v\pdx) - T_\rho - mc^2\right]\varphi_\lambda = mc^2\phi_\lambda.\label{eq:eqphi}
 }
\end{equation}
The couple $(\phi_\lambda,\varphi_\lambda)$ can be formed by either the real frequency modes $(\pom,\vpom)$, or the complex frequency modes $(\xl,\el)$ or $(\psl,\zel)$ and, accordingly, $\lambda$ can be real or complex. Eliminating $\varphi_\lambda$ from the above system, one obtains~\cite{mp}
\begin{equation}\label{eq:modes}
 \left\{\left[\hbar(\lambda +\ii v\pdx) + T_\rho\right] \frac{1}{c^2}\left[-\hbar(\lambda +\ii v\pdx) + T_\rho \right]-\hbar^2 v \pdx\frac{1}{v}\pdx\right\}\phi_\lambda=0.
\end{equation}
When the background quantities are independent of $x$,  Fourier modes $\phi_\lambda\propto\exp(\ii k_\lambda x)$ are solutions of~\eqref{eq:modes}, provided $k_\lambda$ is a (possibly complex) root of the dispersion relation
\begin{equation}\label{eq:dispersion}
 (\lambda-vk)^2=c^2k^2+\frac{\chor^4 k^4}{\Lambda^2}\equiv\Omega^2 
(k),
\end{equation}
where $\Omega$ is the frequency in the comoving frame, and where $\chor$ is a typical value of the sound speed, we take to be that at the horizons, see~\sref{sec:bhlasers}. We also introduced
\begin{equation}
 \Lambda\equiv\frac{2m \chor^2}{\hbar},
\end{equation}
which gives the characteristic dispersive scale.
When sending $\Lambda \to \infty$, \eqr{eq:dispersion} becomes the relativistic equation $\Omega^2 = c^2 k^2$ since the quartic term drops out. Similarly, \eqr{eq:modes} becomes the Euler equation governing sound waves.
Instead, when keeping this term, the dispersion relation \eqr{eq:dispersion} possesses four roots, some of which can be complex. As we shall progressively see, the two extra roots are at the origin of the laser effect.

\subsubsection{Modes in homogeneous condensates.}
\label{sec:modeshom}

In subsonic flows, $|v|< c$, for real $\lambda=\omega$, two roots are real, and describe the right and left moving solutions.
The other two are complex, conjugated to each other, and correspond to modes that are asymptotically growing or decaying, say to left.
Only the first two should be used in \eqr{eq:phiexpansion}, as the last two are not asymptotically bound.

\begin{figure}
 \centering
 \psfrag{om}[c][c]{\small $\Omega/\kappa$}
 \psfrag{k}{\small $\chor k/\kappa$}
 \hspace{0.10\textwidth}
 \includegraphics[width=0.78\textwidth]{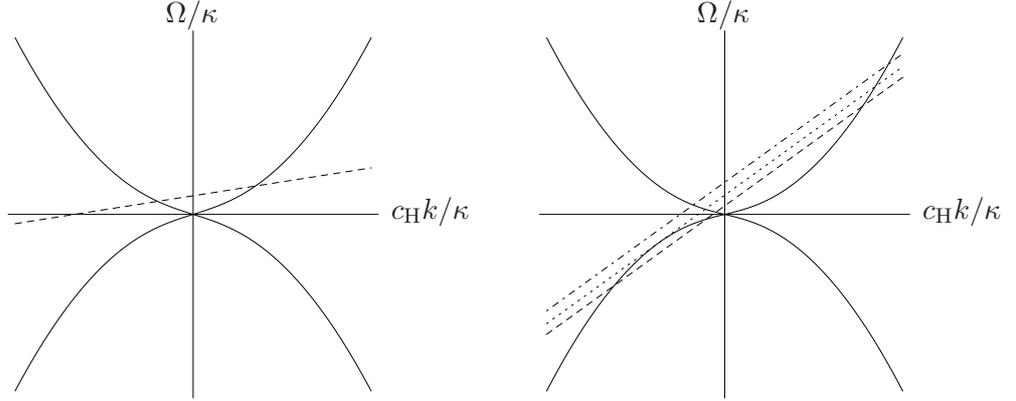}
 \caption{Graphical solution of the dispersion relation~\eref{eq:dispersion} for subsonic flow (left panel) and supersonic flow (right panel). Solid lines: $\pm \Omega(k)/\kappa$. Left panel, dashed line: $(\om-vk)/\kappa$. Right panel, dashed, dotted, dotdashed lines: $\om-vk$ respectively for $\om<\ommax$, $\om=\ommax$, $\om>\ommax$. For $\om<\ommax$, one clearly sees that, in supersonic flows, two extra (real) roots exist in the left lower quadrant.
The most negative one is called, in the text, $k^{(1)}_{\om}$ and the other one $k^{(2)}_{\om}$, so that $k^{(1)}_{\om} < k^{(2)}_{\om} < 0$.}
 \label{fig:dispersion}
\end{figure}

In supersonic flows, the situation is quite different.
For real $\lambda=\omega$, there exists a critical frequency $\om_{\max}$ such that the four roots of \eqr{eq:dispersion} are~\cite{mp}
\begin{itemize}
\item $\om<\ommax$: all real;
 \item $\om>\ommax$: two real and two complex ones, as in subsonic flows.
\end{itemize}
See~\fref{fig:dispersion} for a graphical solution of the dispersion relation \eqr{eq:dispersion}.

When $\om$ is real, the normalization of the (bound) modes can be easily worked out, as it is, up to a trivial factor, independent of the constant velocity $v$ (because of Galilean invariance). Let us write
\begin{eqnarray}
 \pom\ee^{-\ii\om t}=\frac{\ee^{-\ii\om t+\ii k_\om^\alpha x}}{\sqrt{2\pi\rn}}u_\om^\alpha,
 \qquad \vpom\ee^{-\ii\om t}=\frac{\ee^{-\ii\om t+\ii k_\om^\alpha x}}{\sqrt{2\pi\rn}}v_\om^\alpha,
\end{eqnarray}
where $\alpha$ spans over four values if $|v|>c$ and $\om<\ommax$ and over two values elsewhere.
Using the mode normalization~\eqref{eq:normphi}
\begin{equation}\label{eq:uvcond}
 (|u_\om^\alpha|^2-|v_\om^\alpha|^2)\frac{\partial\om}{\partial k}=1,
\end{equation}
from the mode equation~\eqref{eq:eqphi} and the dispersion relation, one has
\begin{equation}\label{eq:dku}
 D_{k_\om^\alpha} u_\om^\alpha = v_\om^\alpha,
\end{equation}
where
\begin{equation}
 D_k=\frac{1}{mc^2}\left[\hbar\sqrt{c^2k^2+\frac{\hbar^2k^4}{4m^2}} - \frac{\hbar^2k^2}{2m} - mc^2\right]
\end{equation}
is independent of $v$.
Equations~\eref{eq:uvcond} and~\eref{eq:dku} fix $u_\om^\alpha$ and $v_\om^\alpha$ except for a common phase.
Taking both $u_\om^\alpha$ and $v_\om^\alpha$ real, one obtains
\begin{equation}\label{eq:normmode}
\eqalign{
 \pom=\sqrt{\frac{\partial k_\om^\alpha}{\partial\om}}\frac{1}{\sqrt{1-D_{k_\om^\alpha}^2}}\frac{\ee^{\ii k_\om^\alpha x}}{\sqrt{2\pi\rn}},\\
 \vpom=\sqrt{\frac{\partial k_\om^\alpha}{\partial\om}}\frac{D_{k_\om^\alpha}}{\sqrt{1-D_{k_\om^\alpha}^2}}\frac{\ee^{\ii k_\om^\alpha x}}{\sqrt{2\pi\rn}}.}
\end{equation}
In supersonic flows, when $0< \om < \ommax$, the two modes associated with the extra real roots of \eqr{eq:dispersion} have negative norm. However, for $-\ommax< \om<0$, the corresponding modes have a positive norm. So, the field $\hp$ is expanded as%
\footnote{%
In terms of the doublets $\hW$ defined in appendix~\hyperlink{app:quantization}{B} one has
\begin{equation}
 \fl\hW=\int_{0}^{\infty}\!\dom\left[W_\om^u\ha_\om^u+W_\om^v\ha_\om^v+\bar W_\om^u\ha_\om^{u\dagger}+\bar W_\om^v\ha_\om^{v\dagger}\right] 
 +\int_{0}^{\ommax}\!\dom\sum_{i=1,2}\left[W_{-\om}^{(i)}\ha_{-\om}^{(i)}+
\bar W_{-\om}^{(i)}\ha_{-\om}^{{(i)}\dagger}\right].
\end{equation}
}
\begin{eqnarray}
  \fl \hp=\int_0^{\infty}\!\dom\!\!\!&&\!\!\left( \ee^{-\ii\om t} \left\{ \left[\phi_\om^u \ha_\om^u+\phi_\om^u\ha_\om^v\right]
+ \theta(\ommax - \om)
\sum_{i=1,2}(\varphi_{-\om}^{(i)})^*\ha_{-\om}^{{(i)}\dagger}\right\} \right.
 \nonumber \\
&&\left. + \ee^{+ \ii\om t} \left\{ \left[(\varphi_\om^u)^*\ha_\om^{u\, \dagger}+(\varphi_\om^u)^*\ha_\om^{v\, \dagger}\right]
+ \theta(\ommax - \om)
\sum_{i=1,2}\phi_{-\om}^{(i)}\ha_{-\om}^{{(i)}}\right\} \right), 
\end{eqnarray}
where $\theta(z)$ is the Heaviside function.

In subsonic flows one obtains the same expression, but without the last term in each bracket since the extra real roots are absent.

\subsubsection{Modes in black-hole--white-hole geometries.}
\label{sec:laser}

To obtain the spectrum in condensates that cross twice the speed of sound, i.e. in flows defining the geometry of \eqr{eq:velocity}, we need to take into account the nontrivial propagation in the non-homogeneous regions localized near the two horizons, and the fact that each subsonic region is now defined on the half line and no longer on the full real axis.
Then, we need to identify the number of globally defined independent modes.

It should be noted that this set does not depend on the particular form of the mode equation, but mainly resides on the behaviour in the internal region II and in the asymptotic flat external regions I and III described in~\sref{sec:bhlasers}.
Therefore, the result is the same when studying a phonon field in a BEC, as in the present paper, or a generic scalar field with superluminal dispersion relation as in~\cite{cp}.
As explained in this reference, in geometries described by~\eqr{eq:velocity}, there is a continuous double set (right-going and left-going waves) of real frequency modes, and the discrete set of complex frequency modes is generally not empty.
Here we give a more mathematical argument yielding the same result.

In flows \eqr{eq:velocity}, two asymptotically flat and subsonic regions play the most important role in defining the set of modes, for both $\lambda$ real and $\lambda$ complex.
For $\lambda= \om$ real, there are three bounded modes: the two usual propagating modes associated with the real roots of \eqr{eq:dispersion} {\it plus} the mode that asymptotically decays. This mode should now be taken into account as it is bound in the subsonic region where it is defined.
We will call the two sets of three modes $\phi_\om^{i,{\rm I}}$ and $\phi_\om^{i,{\rm III}}$, respectively for the left (I) and right (III) asymptotic region, where the superscript $i$ takes three values: $ u, v, CJ$ to characterize the asymptotic right-going $u$-mode, the left-going $v$-mode and the decaying $CJ$-mode, named after the paper of Corley and Jacobson~\cite{CJold}.
In the internal region II, we have four modes that we call $\phi_\om^{j,{\rm II}}$ with $j=1,2,3,4$.
Since this region is bound, the four roots of \eqr{eq:dispersion} should be taken into account.

Therefore, any globally defined solution of frequency $\om$ can be expanded in two forms by referring to its left, or right, asymptotic behaviour:
\begin{equation}
\eqalign{
\phi_\om=\sum_i L_\om^i \, \phi_\om^{i,{\rm I}} ,  \\
\phi_\om=\sum_i R_\om^i \, \phi_\om^{i,{\rm III}}.}
\end{equation}
Moreover, each partial wave $\phi_\om^{i,{\rm I}}$ or $\phi_\om^{i,{\rm III}}$ can be written in term of the $\phi_\om^{j,{\rm II}}$:
\begin{equation}
\eqalign{
\phi_\om^{i,{\rm I}}=\sum_j \mathcal{L}_\om^{ij} \, \phi_\om^{j,{\rm II}}, \\
 \phi_\om^{i,{\rm III}}=\sum_j \mathcal{R}_\om^{ij} \, \phi_\om^{j,{\rm II}}.
 }
\end{equation}
Putting together the above equations, one obtains 
\begin{equation}\label{eq:system}
 \sum_i \mathcal{R}_\om^{ij}R_\om^i=\sum_i \mathcal{L}_\om^{ij}L_\om^i.
\end{equation}
This system of four equations in six unknowns (the coefficients $R_\om^i$ and $L_\om^i$) has a two-dimensional set of solutions, corresponding to two linearly independent modes.
For instance, one can choose as independent solutions, $\phi^{u, in }_{\om}$ and $\phi^{v, in }_{\om}$, the two right-going and left-going in-modes, defined as the solution of~\eqref{eq:system} with respectively $L^u=1,R^v=0$ and $L^u=0,R^v=1$.
Similarly, one can chose the out-modes $\phi^{u, out}_{\om}$ and $\phi^{v, out}_{\om}$ defined respectively by $L^v=0,R^u=1$ and $L^v=1,R^u=0$.
Using the scalar product of \eqr{eq:normphi} to normalize these modes,  
the two sets are related by~\cite{cp}
\begin{eqnarray}
\eqalign{
\phi^{u, in }_{\om} &= T_{\om} \, \phi^{u, out}_{\om} 
+ R_{\om} \,  \phi^{v, out}_{\om}, \\
\phi^{v, in }_{\om} &= \tilde T_{\om}\,  \phi^{v, out}_{\om} + \tilde
R_{\om}\,  \phi^{u, out}_{\om} .  
}
\label{RT}
\end{eqnarray}
Unitarity imposes $| T_{\om}|^2 +|R_{\om}|^2 = 1 = | \tilde T_{\om}|^2 +|\tilde R_{\om}|^2 $, and $R_\om \tilde T_\om^* + T_\om \tilde R_\om^* = 0$.

Let us now move to the complex frequency case. Because all expressions are analytic in $\lambda$, the above analysis applies as such when replacing $\om$ by $\lambda = \om + i \Gamma$.
The novel aspects only come through the condition that the modes be asymptotically bound.
Indeed, $\Gamma >0$ implies that $k^u_\lambda$, the asymptotic wave number of the $u$ real root of \eqr{eq:dispersion} (more precisely of its analytical continuation in $\Gamma$), acquires a positive imaginary part. Thus the mode diverges at $x\to-\infty$, unless one puts $L^u=0$.
Similarly, on the right side, the $v$ mode diverges at $x\to+\infty$, and imposing that it is bound requires $R^v=0$.
However, these two conditions imply that the system~\eqref{eq:system} has only the trivial solution $R^i=L^i=0$, {\it except} when its determinant vanishes. This condition defines an equation for the complex frequency $\lambda$ that has a finite set of solutions: $\{\la, a = 1, 2, ..., N\}$.

\subsection{The semi-classical treatment of~\cite{cp}}
\label{sec:wkb}

Having established the condition that defines the complex frequencies $\la$, we now turn to the question of calculating them.
Since the above algebraic method does not depend on the specific form of the mode equation, the semi-classical treatment of~\cite{cp} applies to the present case.
Therefore, we just summarize the results without going into the details. One should nevertheless pay attention to the fact that one is dealing with modes doublets $(\phi,\varphi)$ and to the normalization of these modes.

When the background is not homogeneous but $v$ varies slowly with respect to the wavelength of the perturbation, the exact solutions are well approximated by their WKB approximation, which can be directly inferred from~\eqref{eq:normmode}
\begin{equation}
\eqalign{
 \pom=\sqrt{\frac{\partial k_\om^\alpha(x)}{\partial\om}}\frac{1}{\sqrt{1-D_{k_\om^\alpha(x)}^2}}\frac{\exp\left[\ii\int^x \dx' k_\om^\alpha(x')\right]}{\sqrt{2\pi\rn(x)}},\\
 \vpom=\sqrt{\frac{\partial k_\om^\alpha(x)}{\partial\om}}\frac{D_{k_\om^\alpha(x)}}{\sqrt{1-D_{k_\om^\alpha(x)}^2}}\frac{\exp\left[\ii\int^x \dx' k_\om^\alpha(x')\right]}{\sqrt{2\pi\rn(x)}}.}
\end{equation}
The $x$-dependent wave numbers $k_\om^\alpha(x)$ are real roots of the dispersion relation \eqr{eq:dispersion} in a stationary inhomogeneous flow characterized by $v(x)$ and $c(x)$:
\begin{equation}\label{eq:dispersion2}
 (\om-v(x)k)^2=c(x)^2 k^2+\frac{\hbar^2 k^4}{4 m^2}. 
\end{equation}
Ignoring the quartic term, one obtains the dispersion relation of a massless field propagating in the acoustic metric \eqr{eq:metric}.
Equations are valid in the three regions I, II, III but not close to the horizons where the gradients of $v$ and $c$ cannot be neglected.

The theoretical analysis can then be greatly simplified when taking into account the fact that the mixing between $u$ and $v$ modes is usually negligible~\cite{mp,mp2}.
(However the numerical analysis of the next section does not rely on this simplifying assumption.)
In this approximation, the laser effect is completely due to the $u$-modes (for flows to the left $v< 0$).
Considering the $u$ in-mode in the internal region II, it can be written as a superposition of three WKB waves:
\begin{equation}\label{eq:expnov}
 \phi_\om^{u,{\rm in}}={\cal A}_\om \phi_\om^u + {\cal B}_\om^{(1)} {\varphi_{-\om}^{(1)}}^* + {\cal B}_\om^{(2)} {\varphi_{-\om}^{(2)}}^*,
\end{equation}
where $\varphi_{-\om}^{(i)}$ are the two negative frequency $\varphi^u$-modes, corresponding to the two extra roots $k_\om^{(i)}$, $\kone$ being the most negative one, see figure~\ref{fig:dispersion}.

In~\cite{ulf} it was shown that the propagation from one horizon to the other is efficiently described by a unitary scattering of a two-component vector, whose first component represents the $u$-wave $\phi_\om^u$ and the other represents the trapped wave, described either by $(\varphi_{-\om}^{(1)})^*$ or $(\varphi_{-\om}^{(2)})^*$.
The $S$ matrix is decomposed in four elementary matrices
\begin{equation}
 S=U_4 U_3 U_2 U_1,
\end{equation}
representing (from 1 to 4) the scattering at the white hole horizon, the propagation between the white and the black hole horizon of $\phi_\om^u$ and $(\varphi_{-\om}^{(1)})^*$, the scattering at the black hole horizon and the propagation back to the white hole horizon of $(\varphi_{-\om}^{(2)})^*$. These matrices are
\begin{equation}\label{eq:umatrices}
 \eqalign{
 U_1=\pmatrix{\alpha_\om & \alpha_\om z_\om \cr \tilde\alpha_\om z_\om^* & \tilde\alpha_\om},\quad
 U_2=\pmatrix{\ee^{\ii S_\om^u} & 0 \cr 0 & \ee^{-\ii S_{-\om}^{(1)}}},\\
 U_3=\pmatrix{\gamma_\om & \gamma_\om w_\om \cr \tilde\gamma_\om w_\om^* & \tilde\gamma_\om},\quad
 U_4=\pmatrix{1 & 0 \cr 0 & \ee^{\ii S_{-\om}^{(2)}}},
 }
\end{equation}
where
\begin{equation}
 S_\om^u\equiv\int_{-L}^L\dx\,k^u_\om(x),\quad
 S_{-\om}^{(i)}\equiv\int_{-L_\om}^{R_\om}\dx\,\left[-k_\om^{(i)}(x)\right],\quad i=1,2,
\end{equation}
are the Hamilton--Jacobi actions governing the phase of the WKB modes, and $L_\om$ and $R_\om$ are the two turning points. Moreover, by unitarity, the parameters in $U_1$ and $U_3$ must satisfy $|\alpha_\om|^2=|\tilde\alpha_\om|^2$, $|\gamma_\om|^2=|\tilde\gamma_\om|^2$, $|\alpha_\om|^2(1-|z_\om|^2)=|\gamma_\om|^2(1-|w_\om|^2)$.

For real frequency modes, the condition that the trapped mode be single valued imposes~\cite{cp}
\begin{equation}
 \spin{\ee^{\ii\theta_\om}}{b_\om}= S\spin{1}{b_\om},
\end{equation}
where $\theta_\om$ is real by unitarity.
From this matricial equation, one obtains
\begin{equation}\label{eq:btheta}
 b_\om= \frac{S_{21}}{1-S_{22}},\qquad \ee^{\ii\theta_\om}=-\frac{S_{11}}{S_{22}^*}\frac{1-S_{22}^*}{1-S_{22}}, 
\end{equation}
and therefore the coefficients of~\eqref{eq:expnov} are
\begin{equation}\label{eq:abb}
 {\cal A_\om}=\alpha_\om(1+z_\om b_\om),\quad {\cal B}_\om^{(1)}=\tilde\alpha(z_\om^*+b_\om),\quad {\cal B}_\om^{(2)}=b_\om.
\end{equation}

For complex frequency with $\Gamma>0$, imposing that the incoming $u$ branch vanishes, one obtains
\begin{equation}
 \spin{\beta_a}{1}=S\spin{0}{1},
\end{equation}
which implies
\begin{equation}\label{eq:beta}
 \beta_a=S_{12},\quad S_{22}=1.
\end{equation}

Equations~\eqref{eq:btheta},~\eqref{eq:abb} and~\eqref{eq:beta} are the central equations we shall use to analyse the results of the numerical integration.
Even though these equations have been obtained through a semi-classical reasoning, their validity goes beyond that of the WKB approximation, as shall be established by the remarkable agreement between the relations one can derive from them and the numerical results.

\begin{figure}
 \centering
 \psfrag{om}[c][c]{\small $\omega/\ommax$}
 \psfrag{b2}[c][c]{\small $|{\cal B}_\om^{(2)}|^2$}
 \includegraphics[width=\textwidth]{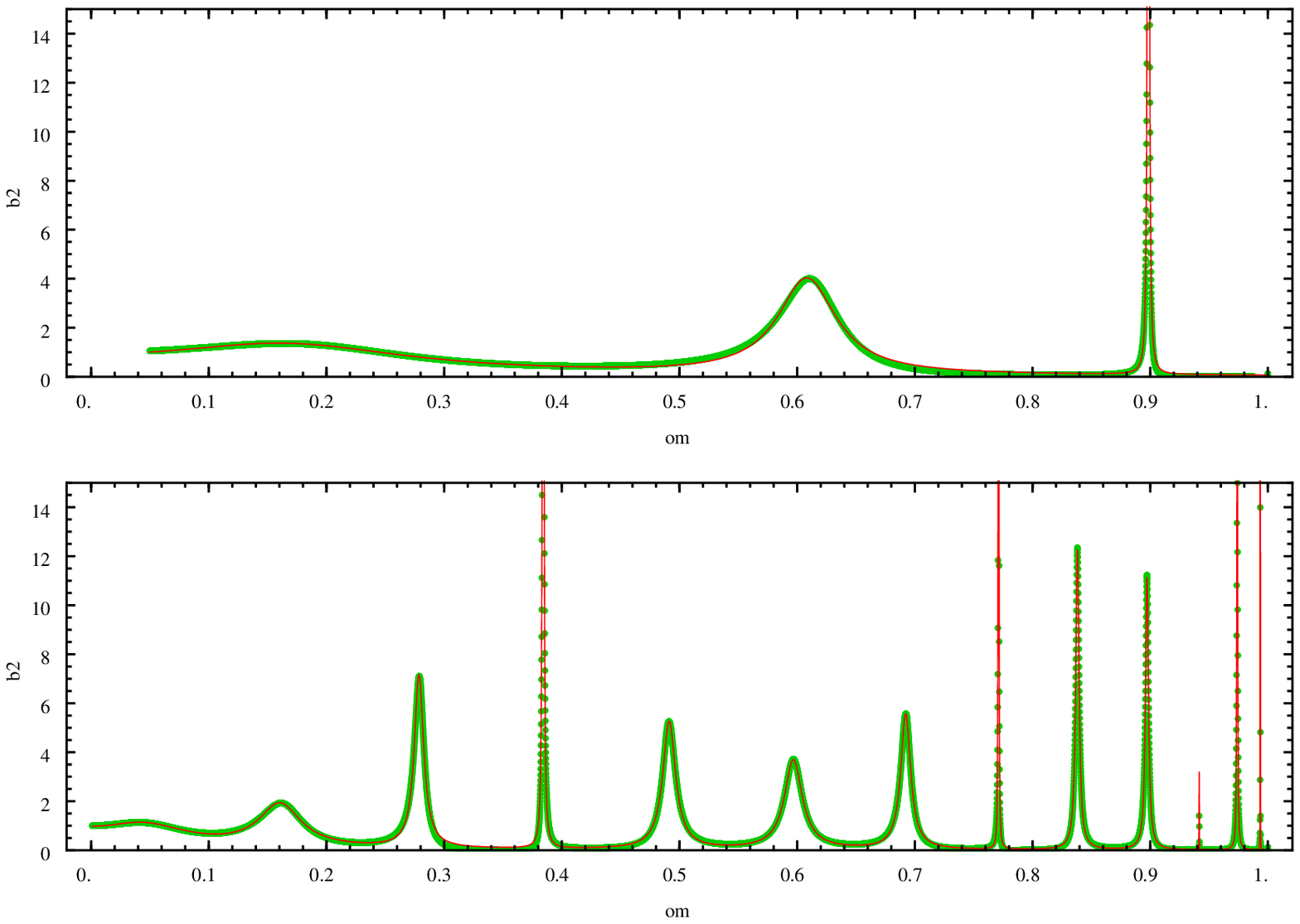}
 \caption{We represent $|{\cal B}_\om^{(2)}|^2$ as a function of $\omega/\ommax$ for two values of $L$, $L\kappa/\chor=6$ (upper panel), $L\kappa/\chor=25$ (lower panel), and for  $\kw=\kb=\kappa$,  $q=0.5$, $D=0.33$, $n=1$, $\Lambda/\kappa=2$, $\ommax/\kappa\approx0.195$. Green points: numerical simulation; red lines: fitted series of Lorentzians. Online movie \href{http://people.sissa.it/finazzi/bec_bhlasers/movies/eigenfrequencies.gif}{eigenfrequencies.gif} (available from~\urlnjp): Evolution of $|{\cal B}_\om^{(2)}|^2$ from $L\kappa/\chor=15$--25.}
 \label{fig:lorentzians}
\end{figure}

The main prediction concerns the relation between $\la$, solutions of ~\eqref{eq:beta}, and the behaviour of the coefficients of \eqr{eq:abb}.
Indeed, $\la$ solution of $S_{22}=1$, is a pole of ${\cal B}_\om^{(2)}=b_\om=S_{21}/(1-S_{22})$. Therefore, when $\Gamma_a$ is small enough (this is always true in our numerical situations), the pole is close to the real axis and the behaviour of ${\cal B}_\om^{(2)}$ for real frequencies $\om$ is dominated by the contribution of the pole:
\begin{equation}
 {\cal B}_\om^{(2)}=\frac{S_{21}}{1-S_{22}}\approx -\frac{\ii\Gamma_a}{\om-(\om_a+\ii\Gamma_a)}{\cal B}_{\om_a}^{(2)},
\end{equation}
from which
\begin{equation}\label{eq:lorentzian}
 |{\cal B}_\om^{(2)}|^2\approx \frac{\Gamma_a^2}{(\om-\om_a)^2+\Gamma_a^2}|{\cal B}_{\om_a}^{(2)}|^2.
\end{equation}
Thus, $|{\cal B}_\om^{(2)}|^2$ should be well described by a sum of Lorentzians characterized by the complex frequencies $\la$.
In~\fref{fig:lorentzians} we plot $|{\cal B}_\om^{(2)}|^2$ obtained from numerical simulations and the corresponding fitted sum of Lorentzians. The agreement is excellent.

We conclude this analysis by showing that a semi-classical treatment furnishes approximate expressions for the complex frequencies $\la$.
Assuming $|z_\om|^2$, $|w_\om^2|$ and $|z_\om w_\om|$ are much smaller than 1, one can expand the condition $S_{22}=1$ in powers of these quantities.
To zeroth order, one has $\lambda_a=\om_a$ real, with $\om_a$ solution of the Bohr--Sommerfeld condition~\cite{cp}
\begin{equation}\label{eq:bs}
 \fl
 S_{-\om_a}^{(1)}-S_{-\om_a}^{(2)}-\arg(\tilde\alpha_{\om_a}\tilde\gamma_{\om_a})
 =\int_{-L_{\om_a}}^{R_{\om_a}}\dx\,\left[-k_{\om_a}^{(1)}(x)+k_{\om_a}^{(2)}(x)\right]
 -\arg(\tilde\alpha_\om\tilde\gamma_\om)=2\pi \nbs,
\end{equation}
where $\nbs \equiv a = 1,2,\ldots,N$.
A semi-classical treatment gives $\arg(\tilde\alpha_\om\tilde\gamma_\om)=-\pi$. A more refined estimate~\cite{cfp} gives
\begin{equation}
\eqalign{
 \arg(\tilde\alpha_\om\tilde\gamma_\om)\approx - \pi - f\left(\frac{\om}{\kb}\right) - f\left(\frac{\om}{\kw}\right),\\
  f(x)\equiv  \arg\Gamma_{\rm E}(\ii x) -  x\log(x) + x +\frac{\pi}{4},\label{eq:f}
  \label{eq:arg}
  }
\end{equation}
where $\Gamma_{\rm E}$ is Euler's Gamma function.

To first order in $|z_\om|^2$, $|w_\om^2|$, $|z_\om w_\om|$ and $\delta\lambda_a=\delta\om_a+\ii\Gamma_a$, $S_{22}=1$ gives
\begin{eqnarray}
 {2 T^{\rm b}_{\om_a}} \Gamma_a = 
\left(|z_{\om_a}|^2+|w_{\om_a}^2|+2|z_{\om_a} w_{\om_a}^*|\cos\vartheta_a\right)=|S_{12}(\om_a)|^2,\label{eq:gamma}\\
{T_{\om_a}^{\rm b}} \delta\om_a = - 
|z_{\om_a} w_{\om_a}^*| \sin\vartheta_a,\label{eq:deltaom}\\
 \vartheta_a=S_{\om_a}^u+ S_{-\om_a}^{(1)}+\arg\frac{z_{\om_a} w_{\om_a}^*\alpha_{\om_a}}{\tilde\alpha_{\om_a}},\\
 T_{\om_a}^{\rm b}=\frac{\partial}{\partial\om}\left[ S_{-\om}^{(2)} - S_{-\om}^{(1)} + \arg\tilde\alpha_{\om}\tilde\gamma_{\om} \right]_{\om=\om_a},\label{eq:tb}
\end{eqnarray}
where $T^{\rm b}_{\om_a}$ is the time for the trapped mode to make a full bounce. The phase $\vartheta_a$ modulates the frequency imaginary part $\Gamma_a$ and the first order correction to its real part $\delta\om_a$, as shall be seen in~\sref{sec:L}.

\subsection{Observables: Phonon fluxes and density--density correlation patterns}
\label{sec:correlations}

Having identified the set of eigenmodes, we now consider observable quantities.
We shall study both the phonon flux emitted by the black-hole--white-hole system, and the non-local density--density correlation pattern, as they illustrate very different aspects of the laser effect.

In what follows, we work in quantum settings because we shall assume that the state at the formation of the supersonic region is vacuum.
In classical settings, the dynamical instability (the laser effect) would be present only if the initial density-perturbation possesses a non-vanishing overlap with some complex frequency mode, see section V.A. in~\cite{cp}.
For a perfectly adiabatic formation of the supersonic flow, the amplitudes associated with these modes would be zero.
However, any small deviation from adiabaticity will trigger the instability, thereby engendering a behaviour very similar to that derived in quantum settings.
In fact, as we shall see, the main difference between the quantum and the classical treatment 
is at early times because in the quantum vacuum all complex frequency modes
contribute to the observables through the spontaneous excitation.

To compute observables, we first need to choose the quantum state.
Because of the instability, the energy~\eref{eq:hamiltonianabc} is unbounded from below, and there is no clear definition of the vacuum. Nevertheless, if the formation  of the supersonic region at time $t_0=0$ is adiabatic, and if the temperature condensate is low enough, i.e. much smaller than $\kappa$, the Heisenberg state of the system can be approximated by the state annihilated, at that initial time, by the destruction operators $\aom$, $\dlpl$, $\dlmi$ of \eqr{eq:phiadd}.%
\footnote{We applied the standard reasoning to both the $\aom$ and the $\hat d$ operators because $\Gamma_a$, the imaginary part of the complex frequency $\la$, is smaller than a tenth of $\om_a = \re \la$.}
Since the operators $\dlpl$, $\dlmi$ are not stationary, the expectation values will not be stationary either, and will instead depend on the lapse of time since the formation at $t_0=0$. In what follows all expectation values will be computed in this state.

The two observables we shall study are related to the density fluctuation $\hr=\hat\rho-\rn$.
Given the definition of $\hp$ in~\eref{eq:defphi}, $\hr$ is given by
\begin{equation}
 \hr=\rn(\hp+\hpd)= \rn \, \hch,
\end{equation}
where $\hch\equiv\hp+\hpd$ is Hermitian. Its expansion in terms of operators $\hat a,\hat d$ is
\begin{eqnarray}\label{eq:chiexpansion}
 \fl\hch(t,x)=\int\!\!\dom\sum_{\alpha}\!\left[\phm{\om t}\com\aom+\mbox{h.c.} 
\right]
 +\sum_a\!\left[\clpl\dlpl+\clmi\dlmi+ \mbox{h.c.} 
\right],
\end{eqnarray}
where
\begin{eqnarray}
\label{chimodes}
 \com\equiv\pom+\vpom,\nonumber \\
\chi_{a, \pm}(t,x)\equiv \phi_{a, \pm}(t, x)+ \varphi_{a, \pm}(t, x).
\end{eqnarray}
In the vacuum defined at $t_0=0$, the two-point function is
\begin{eqnarray}%
 \fl \vev{
\hch(t,x) \, \hch(t',x')} &&= 
\int\!\dom\sum_{\alpha}\phm{\om(t-t')}\com\comsxp
\nonumber\\
 \fl && \quad \quad 
 +\sum_a\Big(\clpl\clplsxp+\clmi\clmisxp\Big). 
\end{eqnarray}
Because of the complex frequency modes, it is not a function of $t-t'$ only. This differs from the cases
of a single black hole (or white hole) horizon, where the two-point function is stationary in vacuum~\cite{mp}.

To extract more physical information, it is convenient to return to the frequency eigenmodes appearing in \eqr{eq:phiexpansion}. Defining
\begin{equation}
 \sil\equiv\xl+\el,\quad\nl\equiv\psl+\zel,
\end{equation}
the two-point function becomes
\begin{eqnarray}
\fl 
\vev{
\hch(t,x) \, \hch(t',x')}
= \int\!\dom\sum_{\alpha}\phm{\om(t-t')}\com\comsxp
 \nonumber\\
  +\sum_a\re\left[\ee^{\Gamma(t+t')}\phm{\om_a(t-t')}\sil\silsxp+\ee^{-\Gamma(t+t')}\phm{\om_a(t-t')}\nl\nlsxp\right]
 \nonumber\\
 +\ii\sum_a\re\left[\phm{\la(t-t')}\sil\nlsxp-\phm{\las(t-t')}\nl\silsxp\right].
\end{eqnarray}
The first line gives the (vacuum) contribution of the real frequency modes. Since these are only elastically scattered, see \eqr{RT}, they stay in their ground state and contribute neither to the emitted fluxes nor to the pattern of density--density correlations. The third line contains a mixed contribution of the growing $\sil$ and decaying modes $\nl$. It does not contribute to the fluxes at any time because $\sil$ vanishes in region I, and $\nl$ does it in III, see the paragraph after \eqr{RT}. Moreover, since it only depends on $t-t'$, it does not significantly contribute to the correlation pattern at late time.
Therefore, both fluxes and correlation patterns are governed by the growing mode contribution, the first term  of the second line in $\sil\silsxp$.

It is clear that at late times, i.e. $t-t_0 > 1/{\max}(\Gamma_{a})$, all observables are dominated by the mode with the largest $\Gamma_a$.
For instance, the equal time density--density correlation function is asymptotically given by
\begin{equation}\label{eq:densitycorrelation}
 \vev{\hr(t,x)\hr(t,x')} \sim \rnx\rnxp\ \times \ee^{2\Gamma_a t} \, \re\left[ \sil\silsxp \right].
\end{equation}
The real part of the frequency $\om_a$ drops out and, as a result, the locus of the maxima does not propagate with time.
This function is studied for various modes in~\sref{sec:modesandcorrelation}.
In case the initial state is not vacuum but a thermal state, the above two-point function would be multiplied by $2n_a+1$, where $n_a$ is the mean occupation number, which depends in a nontrivial way both on the temperature and on the blue shift effect due to the horizons (see equation (46) in~\cite{mp}).
Similar considerations apply to the quantity that is considered below.
Had we worked in classical settings, we would have obtained a coherent wave whose behaviour in $x$ and $t$ is the same as that of $\vev{\hr(t,x)\hr(t',x')}$ at fixed $x',t'$, see equation (54) in~\cite{cp}. The main difference arises from the fact that the classical wave amplitude is fixed by initial conditions.

It is also interesting to study the onset of the instability for earlier time, i.e. $t-t_0 < 1/{\max}(\Gamma_{a})$.
To this end, we consider the following quantity:
\begin{eqnarray}
P_{\rm bh-wh}(\om, T) &\equiv& 
\int_{0}^T dt'  \int_{0}^T dt  \,\ee^{-i\om(t-t')} \,  
\vev{ \hch(t,x)\,  \hch(t',x)} ,
\nonumber \\
\nonumber \\
& \sim & 
\Sigma_a \, \ee^{\Gamma_a T}\, \vert \sil 
\vert^2 \, \frac{ 4 \vert \sin[(\om - 
\om_a - i \Gamma_a
)T/2] \vert^2}
{(\om - \om_a)^2 + (\Gamma_a)^2}.
\label{rate}
\end{eqnarray}
It governs the probability that a density fluctuation of frequency $\om$ be observed during the interval $[0,T]$ at some fixed location $x$ in the subsonic region III, see~\cite{cp}.

When $\Gamma_a \to 0$, when the discrete set labeled $a$ becomes continuous, and when $T$ is sufficiently large, $P_{\rm bh-wh}(\om, T)$ behaves as in the Golden Rule: $P_{\rm bh-wh}(\om, T)\propto  T\, \bar n_{\om}$, i.e. the lapse of time $T$ times $\bar n_{\om}$, the mean occupation number of frequency $\om$.
There can be pre-factors that depend on the strength of the coupling, the amplitude of the mode $\phi_\om$ at $x$, if one deals with a derivative coupling.

It is more interesting to study $P_{\rm bh-wh}(\om, T)$ as a function of $T$ for a given black-hole--white-hole geometry, and to compare it with $P_{\rm bh}(\om, T)$, the corresponding quantity computed in the case where only the black hole horizon would be present. When $T \times {\max}(\Gamma_a) \ll 1$, the two observables behave in a very similar manner, even when only a few (say 10) complex frequency modes exist in the black-hole--white-hole case. This shall be seen by a numerical analysis in~\sref{sec:growthofn}, but can be also understood from the properties of the growing modes $\sil$, see the central paragraph of section V.B.5 in~\cite{cp}.

For classical settings, the laser effect is present only when the initial density-perturbation profile is characterized by a non-vanishing amplitude of some complex frequency mode. Since for adiabatic formation of the horizons all the amplitudes associated with those modes are zero, no laser effect appears in this case. However, even a small deviation from perfect adiabaticity causes the presence of instabilities. The main difference in the quantum framework is therefore the possibility to make the system unstable through the spontaneous excitation of complex frequency modes.

\section{Numerical results}
\label{sec:numerics}

\subsection{The method} 
\label{sec:nmethod}

In a few words, we describe how we proceeded to solve numerically \eqr{eq:eqphi} in the flows described by \eqr{eq:velocity} and \eqr{eq:cv}.
As in the case of a single black hole or white hole horizon, when working with a fixed frequency, the main task is to avoid the growing mode contaminating the oscillatory modes.
The way to get rid of this difficulty is by integrating from the subsonic region towards the horizon into the supersonic region where there is no growing mode~\cite{CJold}.

Basically we solved separately the mode equation from region I to region II, and from region III to region II, using a code adapted from~\cite{mp}. In each case, as in that reference, we integrated the equations for initial conditions describing the three acceptable modes in regions I and III: the right-moving $u$-mode, the left-moving $v$-mode and the decaying mode.
Then for each of them we computed the amplitudes of the four oscillatory modes in the supersonic region II. The two globally defined modes are built using the procedure described in~\sref{sec:laser}.
To obtain the scattering from I to III, we eliminated the amplitudes in region II.

To obtain the correlation patterns of figures~\ref{fig:correlations}--\ref{fig:correlations2} we solved the mode equation with the corresponding complex frequency $\la$ that we had formerly computed.


\subsection{The discrete set of complex frequencies}
\label{sec:poles}

Two strategies can be used to determine the complex frequencies $\la = \om_a + i\Gamma_a$, solutions of \eqr{eq:beta}.
They can be obtained either by solving directly the linear system~\eref{eq:system}, or by determining the center and the width of the Lorentzians in $|{\cal B}_\om^{(2)}|^2$ of \eqr{eq:lorentzian}.
We use the latter to localize them and the former to refine the results and study how they depend on the various parameters of the system.

To give a first idea, in~\fref{fig:lorentzians} we represent $|{\cal B}_\om^{(2)}|^2$ as a function of $\omega/\ommax$ (green points) for two different values of $L$, the distance between the horizons. A sum of Lorentzians functions (red line) is fitted to the data obtained from the numerical analysis (see~\eref{eq:lorentzian}).
The quality of the fit confirms the correctness of the theoretical analysis of \sref{sec:wkb} and of~\cite{cp}.
In the following sections, we study the dependence of $\la$ on both the parameters describing the geometry (see the velocity profile~\eref{eq:velocity}) and the dispersive scale $\Lambda$ of~\eref{eq:dispersion}. The dependence on $n$ is not reported here, since it induces no significant change.%
\footnote{
The role of $n$ of \eqr{eq:velocity} is to govern the smoothness of the transition from the near-horizon region to the flat asymptotic ones.
A smaller $n$ corresponds to smoother transition, while a larger $n$ corresponds to steeper transition and, consequently, to a larger almost-flat region between the horizons. The usefulness of $n$ is essentially technical: it allows one to control the numerical analysis when $L$, the distance between the horizons, is comparable to $D$, the width of the transition  between regions I--III of~\sref{sec:bhlasers}.}
Before proceeding we make some comment about the properties and the validity of the Bohr--Sommerfeld condition~\eref{eq:bs}.

\subsubsection{The validity of the semi-classical approximation.}
\label{sec:Validity}

The action appearing in \eref{eq:bs} contains the {\it difference} of two wave vectors: $\ktwo -\kone $.
These have the same sign (negative for positive $\omega$), as they belong to the same $u$-branch of the dispersion relation (see~\fref{fig:dispersion}), but they have opposite group velocity: $\kone$ describes a right-going mode with respect to the lab, whereas $\ktwo$ a left-moving one.
This peculiarity give rise to an unusual phenomenon.
In the usual case, eigenmodes with small (large) frequency correspond to small (large) Bohr--Sommerfeld numbers $\nbs$, i.e to modes with few (many) nodes. In the present case instead, a large $\nbs$ corresponds to low frequency modes and vice versa.
This can be understood from~\eref{eq:bs}.
Because the wave vectors appeared subtracted, a small $\nbs$ implies $\kone$ close to $\ktwo$ and this happens for $\omega$ close to $\ommax$ (see~\fref{fig:dispersion}). On the contrary, a large $\nbs$ implies a large difference between $\kone$ and $\ktwo$, that is, $\omega$ close to 0.

\begin{figure}
 \centering
 \psfrag{om}[c][c]{\small $\omega/\ommax$}
 \psfrag{b2}[c][c]{\small $|{\cal B}_\om^{(2)}|^2$}
 \includegraphics[width=\textwidth]{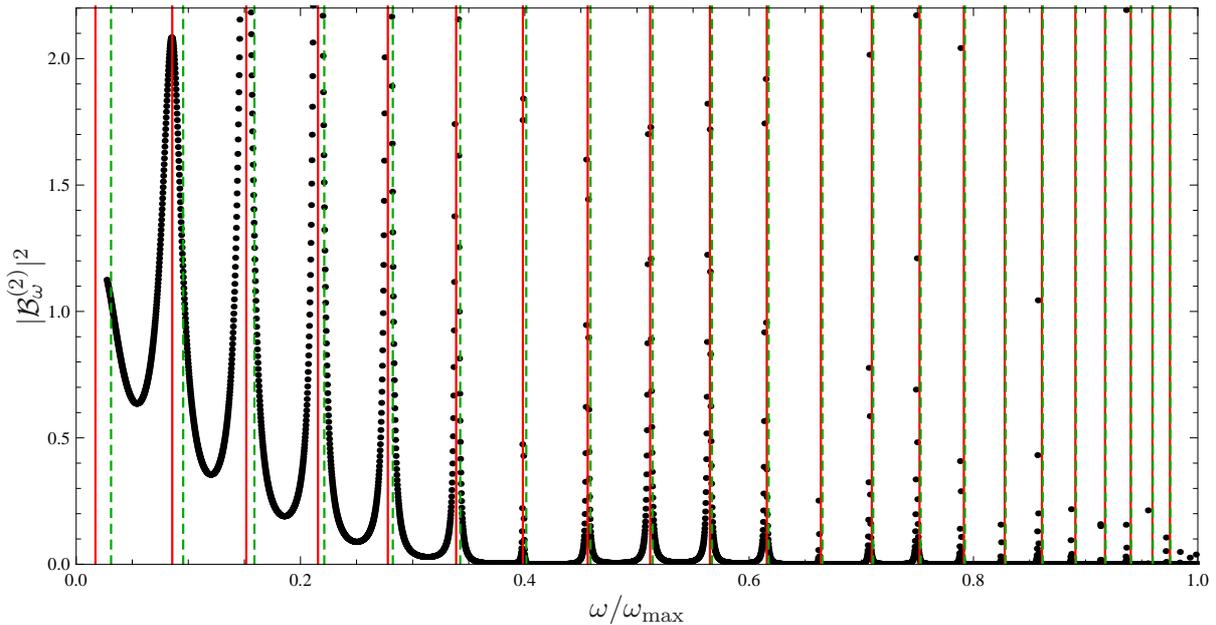}
 \caption{$|{\cal B}_\om^{(2)}|^2$ as a function of $\omega/\ommax$ for $\kw=\kb=\kappa$, $q=0.5$, $D=0.5$, $n=2$, $L\kappa/\chor=10$, $\Lambda/\kappa=8$, $\ommax/\kappa\approx1.408$. Points: numerical simulation; green dashed spikes: standard WKB approximation; red solid spikes: improved treatment of \eqr{eq:arg}.}
 \label{fig:peakswkb}
\end{figure}

This has an unusual consequence. The Bohr--Sommerfeld condition is more reliable when the action is large, that is for $\nbs\gg1$. In the present case this happens for small $\omega/\ommax$. On the other hand the WKB approximation is expected to fail for small $\omega$, as can be verified by the fact that $f$ of \eqr{eq:f} cannot be neglected when $\omega/\kappa \leq 1$.
Both these expectations are confirmed in~\fref{fig:peakswkb} where we compare the numerical results with the predictions obtained with the standard WKB approximation (green lines) and with the improved method (red lines), that is when using \eqr{eq:arg}.
Firstly, the agreement between the Bohr--Sommerfeld condition and the numerical results is worse for $\om$ close to $\ommax$ and gets better when $\nbs$ increases.
Secondly, at low frequency, while the quality of the standard WKB prediction (green lines) becomes worse, the improved method continues to work very well, thereby establishing its validity.
Further studies about the agreement of the improved method and numerical results are in preparation~\cite{cfp}.

%

\subsubsection{The birth of modes as a function of the distance between the horizons.}
\label{sec:L}

\begin{figure}
 \centering
 \psfrag{oma}[c][c]{\scriptsize $\om_a/\ommax$}
 \psfrag{gamma}[c][c]{\scriptsize $\Gamma_a/\kappa$}
 \psfrag{L}[c][c]{\scriptsize $L\kappa/\chor$}
 \includegraphics[width=0.48\textwidth]{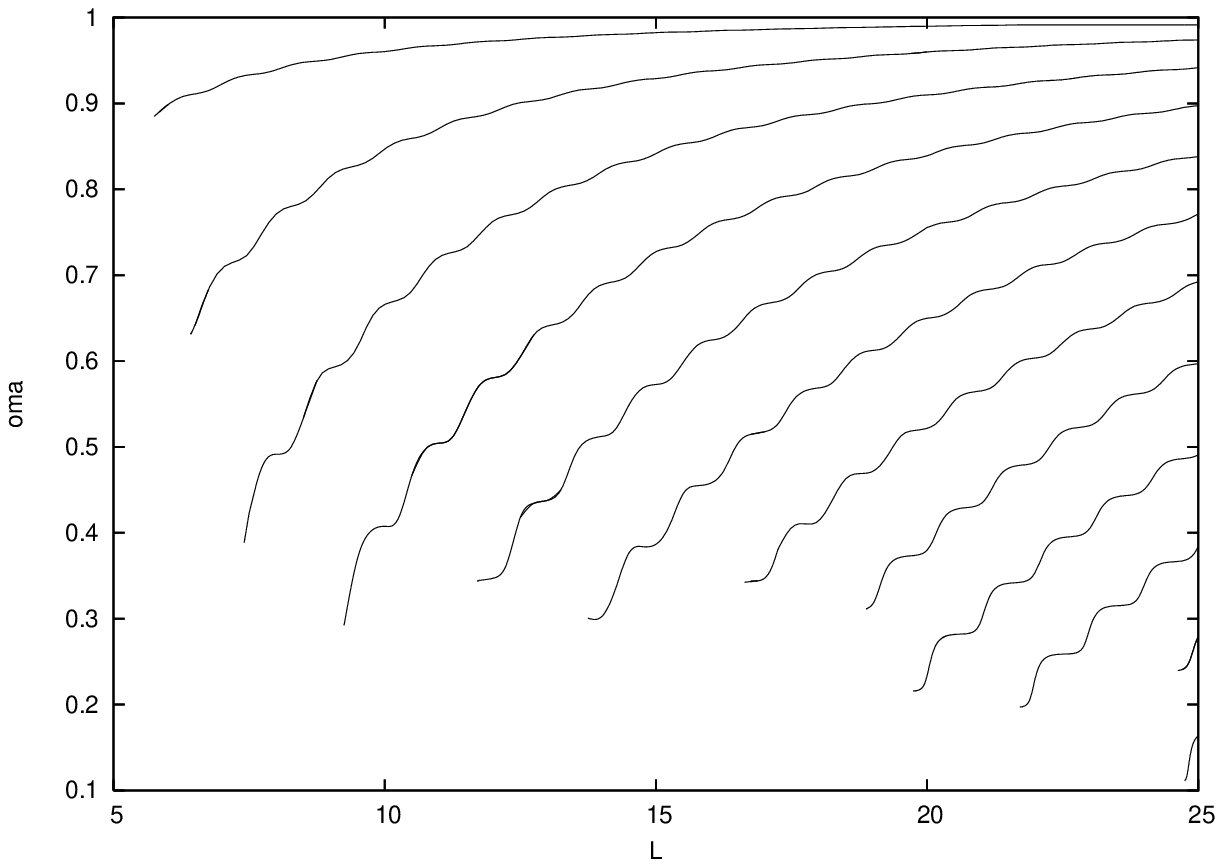}
 \includegraphics[width=0.48\textwidth]{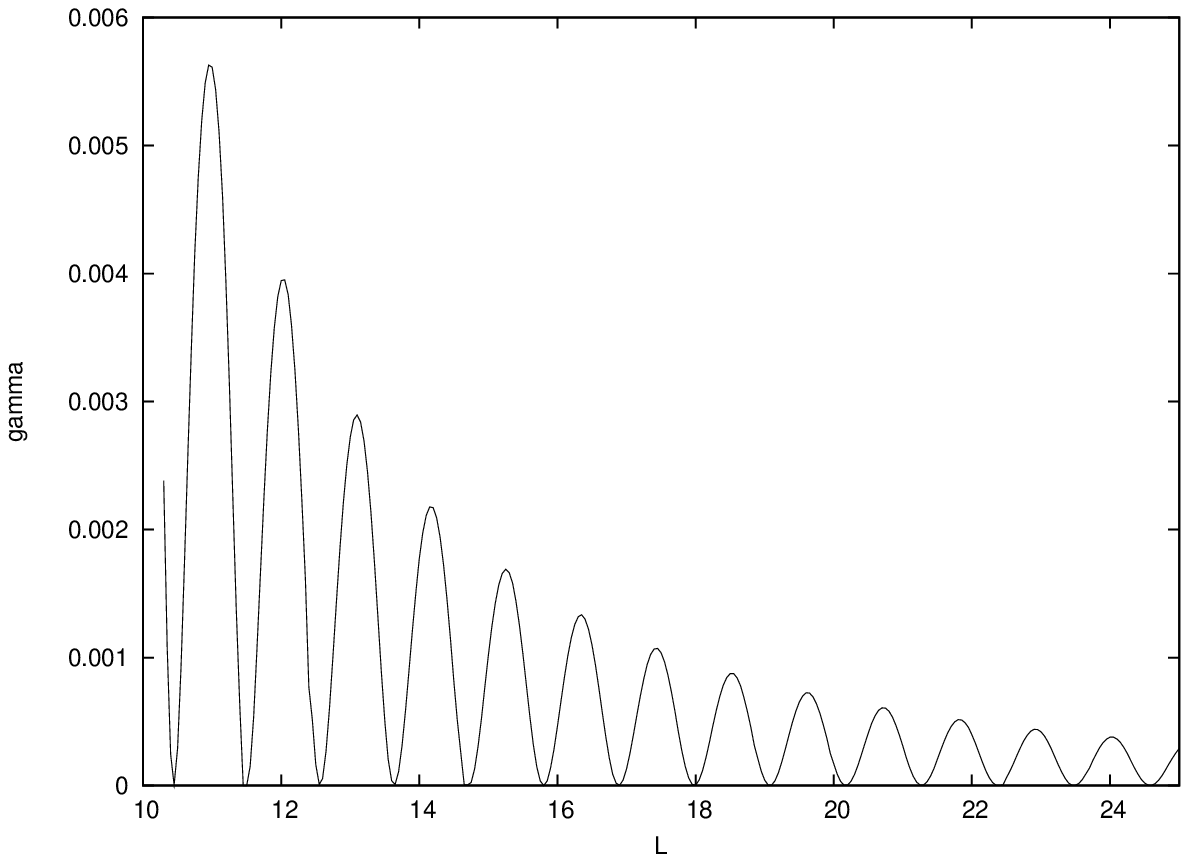}
 \caption{Left panel: The set of $\omega_a/\ommax$ as a function of $L$, from $\nbs$=1 (upper curve) to $\nbs=12$ (lower curve) for the same fixed parameters as in \fref{fig:lorentzians}. At $\kappa L/\chor =25$, the values of $\omega_a/\ommax$ correspond to the peaks of the lower panel of that figure from right ($\nbs=1$) to left ($\nbs=12$).
The broad peak of \fref{fig:lorentzians} near $\om/\ommax =0.05$ cannot be reproduced with this method because of numerical errors. For similar reasons, each curve $\om_a(L)$ ends for low $\omega$. Right panel: $\Gamma_a/\kappa$ for $\nbs=4$ as a function of $L$ for the same fixed parameters. The occurrence of zeros is due to the fact that $\kappa_B=\kappa_W$, as discussed in the text.}
 \label{fig:omega}
\end{figure}

In~\fref{fig:omega}, left panel, we plot $\om_a = \re(\la)$ for the entire spectrum of complex frequency modes as a function of the half-distance between the horizons $L$, all other quantities being fixed.
On the right panel, $\Gamma_a = \im(\lambda_a)$ is plotted as a function of $L$, for $\nbs$=4 and for the same values of the other parameters.
As conjectured in~\cite{cp}, when $L$ grows, all $\om_a(L)$ values increase, and when there is enough `room', a new eigenmode appears with $\omega_a \sim 0$. As a result, the eigenfrequencies become denser close to $\ommax$, because they neither cross it nor disappear.

It is interesting to consider more closely the birth of new modes near $\om=0$.
A rather difficult question to answer is the following: when extra mode appear, does the imaginary part of the frequency $\Gamma_a$ vanish, as found in figure~1 of \cite{Cardoso}, or not?
This question is physically relevant in that it governs the continuous character of the stability of the system: if $\Gamma_a$ vanishes, it implies that the birth of new modes leads to a continuous behaviour in $L$ (as one can expect on the general ground that the properties of the eigenmodes, solutions of \eqr{eq:hp}, continuously depend on the parameters appearing in that equation).
From~\fref{fig:omega}, which is based on the roots of the determinant in~\eref{eq:system}, no definite conclusion can be drawn since there is a loss of numerical precision for $\om \to 0$.
A more reliable method consists in studying the behaviour of $|{\cal B}_\om^{(2)}|^2$ on the real frequency axis. We created a gif-animation (\href{http://people.sissa.it/finazzi/bec_bhlasers/movies/eigenfrequencies.gif}{eigenfrequencies.gif} (available from~\urlnjp)) of $|{\cal B}_\om^{(2)}|^2$ as a function of $\om$ for increasing values of $L$ to illustrate the appearance of the new eigenfrequencies.
From this it is quite clear that when a new eigenmode appears, both the real part and the imaginary part of $\la$ vanish, as was also found in the appendix of \cite{Fullingbook}.

Another important feature confirming the theoretical analysis of~\sref{sec:wkb} is the presence of superposed oscillations on the overall trend of $\omega_a$. In fact, the latter is given by the solution of the Bohr--Sommerfeld equation~\eqref{eq:bs}, whereas the oscillations are due to the deviations governed by the sine in $\delta\om_a$ of~\eref{eq:deltaom}.
For the imaginary part instead, there is no zeroth-order contribution, and the dominant contribution to the right plot of~\fref{fig:omega} is given in~\eref{eq:gamma}.

\subsubsection{Asymmetric velocity profiles.}
\label{sec:kw}

\begin{figure}
 \centering
 \psfrag{oma}[c][c]{\small $\om/\kb$}
 \psfrag{gamma}[c][c]{\small $\Gamma_a/\kb$}
 \psfrag{L}[c][c]{\small $L\kb/\chor$}
  \includegraphics[width=0.6\textwidth]{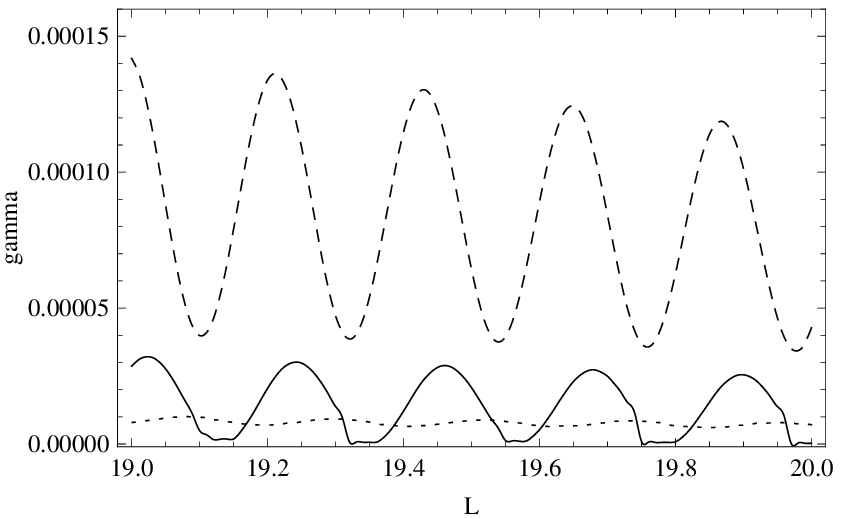}
 \caption{Left panel: $\Gamma_a/\kb$ as a function of $L\kb/\chor$ for $\kw=\kb$ (solid line), $\kw/\kb=0.5$ (dotted line) and $\kw/\kb=2$ (dashed line), for $\ommax/\kb=0.974$, $D=0.33$, $n=1$, $q=0.5$, $\Lambda/\kb=10$.
}
 \label{fig:kw}
\end{figure}

Using asymmetric velocity profiles, namely different surface gravities $\kw$ and $\kb$, does not lead to substantial modifications of the spectrum. 
The only observable change concerns $\Gamma_a$.
When $\kw=\kb$, $\Gamma_a$ vanishes for some particular combination of parameters, as can be seen in~\fref{fig:omega}, right panel. \Eref{eq:gamma} can indeed be rewritten as
\begin{equation}
 \Gamma_a = \frac{1}{2 T_{\om_a}}\left[\left(|z_{\om_a}|-|w_{\om_a}|\right)^2+4|z_{\om_a} w_{\om_a}^*|\cos^2\left(\frac{\vartheta_a}{2}\right)\right],
\label{eq:Ga2}
\end{equation}
which shows that when $|z_{\om_a}|=|w_{\om_a}|$, $\Gamma_a$ vanishes when the cosine does, and this is because the scatterings at the black and the white horizons destructively interfere with each other.
When the two surface gravities are different, $\left(|z_{\om_a}|-|w_{\om_a}|\right)^2$ is non-zero and $\Gamma_a$ can no longer vanish (see~\fref{fig:kw}).
The non-zero offset is significant only for sufficiently large values of $\omega/\kappa$, which requires that $\ommax/\kappa$ be large enough, since all $\om_a < \ommax$.
Using the standard expressions $z_\om= e^{- \pi \om/\kappa_W}$ and $w_\om= e^{- \pi \om/\kappa_B}$, which furnish reliable estimates~\cite{mp}, the ratio between the offset and the amplitude of the oscillation is
\begin{equation}\label{eq:ratio}
 \frac{\left(|z_{\om_a}|-|w_{\om_a}|\right)^2}{4|z_{\om_a} w_{\om_a}^*|}= \sinh^2(\pi \om_a(\kappa_B-\kappa_W)/2).
\end{equation}
Even though this function increases in  $\omega$ when $\kw\neq\kb$, the consequences of this are never important because when the ratio is large, the corresponding mode will not significantly contribute to the instability since $\Gamma_a \ll 1$, as can be seen from \eqr{eq:Ga2}.

\subsubsection{The $u$--$v$ mixing and the parameter $q$.}
\label{sec:Tq}

\begin{figure}
 \centering
 \psfrag{q}[c][c]{\small $q$}
 \psfrag{T2}[c][c]{\small $|T|^2$}
  \includegraphics[width=0.6\textwidth]{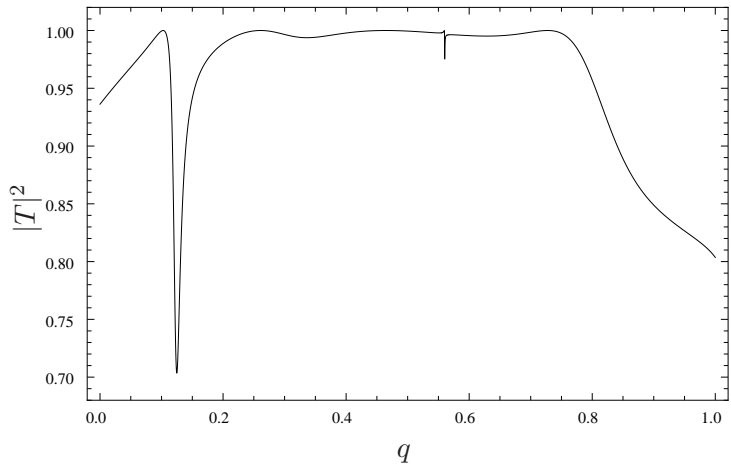}
 \caption{Transmission coefficient $|T_\om|^2$ 
as a function of $q$, at constant $\omega/\kappa=0.100$, and for the other parameters as in the 
bottom plot of  \fref{fig:lorentzians}.
The sharp minima at $q=0.1248$ and $q=0.56023$ correspond to complex frequencies at $\om_a/\kappa=0.100$, whereas the broad ones at $q=0.3363$ and $q=0.6285$ correspond to frequencies respectively at $\om_a/\kappa=0.090$ and $\om_a/\kappa=0.103$.}
 \label{fig:Tq}
\end{figure}

The mixing between the $u$-modes, which are right-going in the lab in subsonic flows, and the $v$-modes is controlled by the transmission coefficient $T_\om$ of \eqr{RT}.
In~\fref{fig:Tq}, $|T_\om|^2$ is plotted as a function of $q$ for a fixed frequency $\om/\kappa=0.100$. The transmission coefficient is almost 1 between $q=0.25$ and $q=0.75$. Thus, for values of $q$ in this range the $u$--$v$ mixing is negligible, as was noticed in~\cite{mp} for a single black (or white) hole.
In this regime the approximation discussed in~\sref{sec:wkb} is valid.
There is, however, an important difference with respect to the single black hole case.
In black-hole--white-hole geometries, the transmission coefficient deviates from zero when approaching a resonance.
When working with a fixed $\omega$, we found two sharp spikes~\fref{fig:Tq} for $q=0.1248$ and $q=0.56023$ which exactly correspond to two unstable modes.
Moreover, the width of the spikes is almost equal to the corresponding $\Gamma_a$. We also looked for complex eigenfrequency associated with the other two local and broad minima of $|T|^2$, at $q=0.3363$ and $q=0.6285$, and we found two eigenmodes with respectively $\om_a/\kappa=0.090$ and $\om_a/\kappa=0.103$, i.e. in the neighbourhood of the chosen frequency $\om/\kappa=0.100$.
These considerations show that close to resonances one cannot completely neglect the $u$--$v$ mixing.
Nevertheless, as shown in~\fref{fig:peakswkb}, the improved treatment of the Bohr--Sommerfeld equation produces very good estimates for the values of the real part $\omega_a$ of the complex frequencies.

\subsubsection{The role of the maximal frequency $\ommax$.}
\label{sec:ld}

\begin{figure}
 \centering
 \psfrag{oma}[c][c]{\small $\om/\ommax$}
 \psfrag{b2}[c][c]{\small $|{\cal B}_\om^{(2)}|^2$}
 \psfrag{L}[c][c]{\small $L\kb/\chor$}
 \psfrag{r}[c][c]{}
  \includegraphics[width=\textwidth]{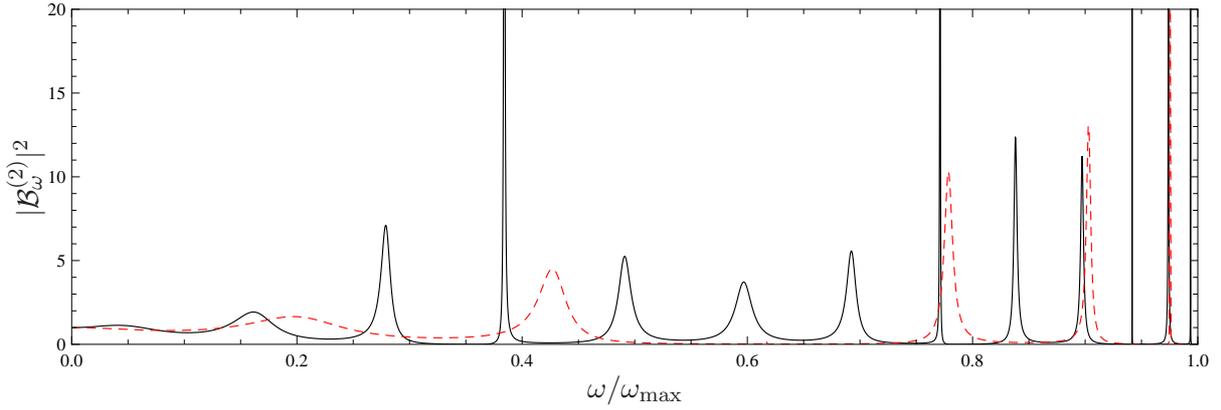}
 \caption{$|{\cal B}_\om^{(2)}|^2$ as a function of $\omega/\ommax$ for different values of $D$ and $\Lambda$ giving rise to the same $\ommax/\kappa=0.19$: black solid line for $D=0.33$, $\Lambda/\kappa=2$, and  red dashed line for $D=0.7$, $\Lambda/\kappa=0.695$.
The other parameters are those of the bottom plot of  \fref{fig:lorentzians}.}
 \label{fig:ld}
\end{figure}

In \cite{mp,mp2}, it was shown that the deviations, due to dispersion and w.r.t. the standard Planckian distribution, of the spectrum  emitted by a single BH, or WH, are mainly governed by $\ommax$. However, the latter depends both on the UV scale $\Lambda$ and the velocity profile (see~\sref{sec:laser}). The relationship takes the form
\begin{equation}
 \ommax=\Lambda f(D,q).
\end{equation}
Hence the same value of $\ommax$ can be reached from very different cases, and yet it was found that the fluxes are hardly sensitive to this. In the present case however, this insensitivity is lost because the number of resonances directly depends on $\Lambda$. This can be understood by studying~\eref{eq:bs}, and is manifest in~\fref{fig:ld}, where $|{\cal B}_\om^{(2)}|^2$ is plotted as a function of $\omega/\kappa$ for two different values of $\Lambda$.

%

\subsection{Growth of the asymptotic phonon fluxes}
\label{sec:growthofn}

\begin{figure}
 \centering
 \psfrag{om}[c][c]{\scriptsize $\om/\ommax$}
 \psfrag{t}[c][c]{\scriptsize $T_{\om_a}^{\rm b}$}
 \psfrag{gamma}[c][c]{\scriptsize $\Gamma_a/\kappa$}
 \psfrag{n}[c][c]{\scriptsize $\dd P/\dd T$}
 \includegraphics[width=.5\textwidth]{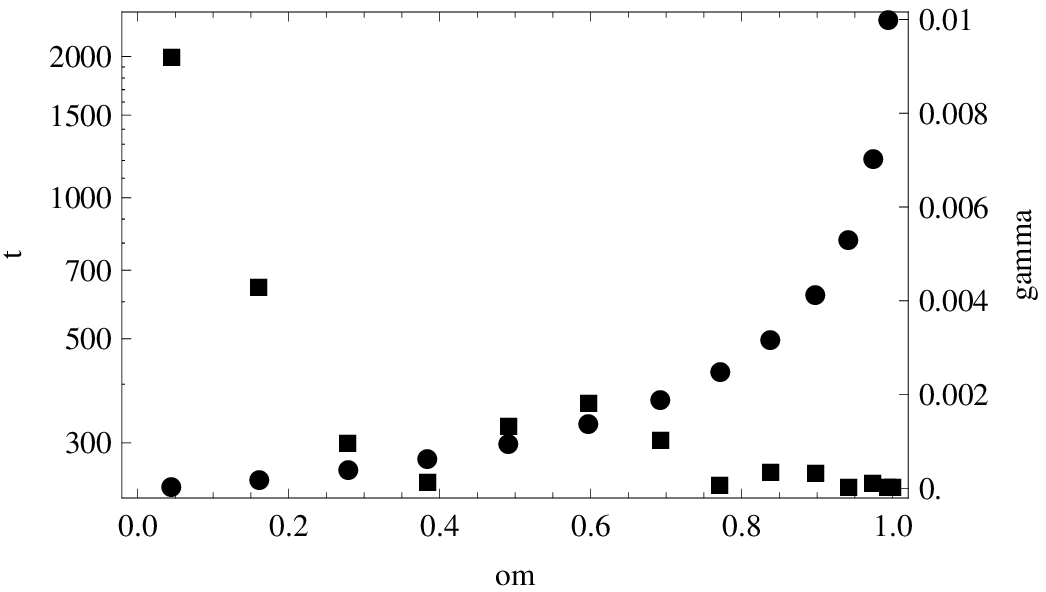}
 \hspace{.01\textwidth}
 \includegraphics[width=.45\textwidth]{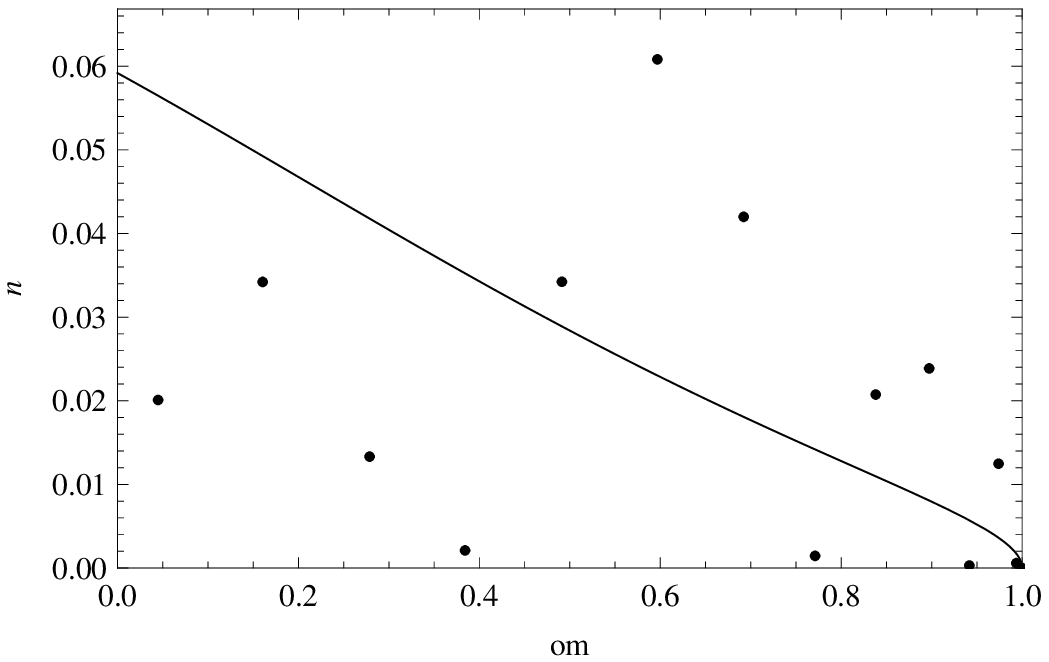}\\
 \psfrag{pr}[c][c]{\scriptsize $P(\om,T=30\kappa^{-1})$}
 \includegraphics[width=.45\textwidth]{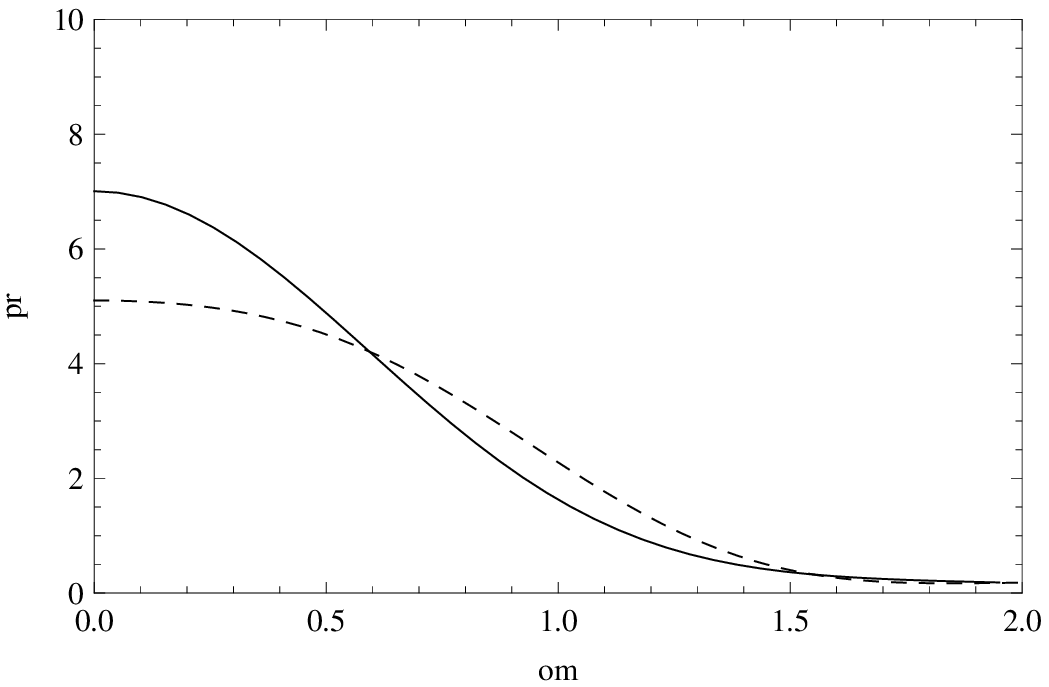}
 \hspace{.06\textwidth}
 \psfrag{pr}[c][c]{\scriptsize $P(\om,T=200\kappa^{-1})$}
 \includegraphics[width=.45\textwidth]{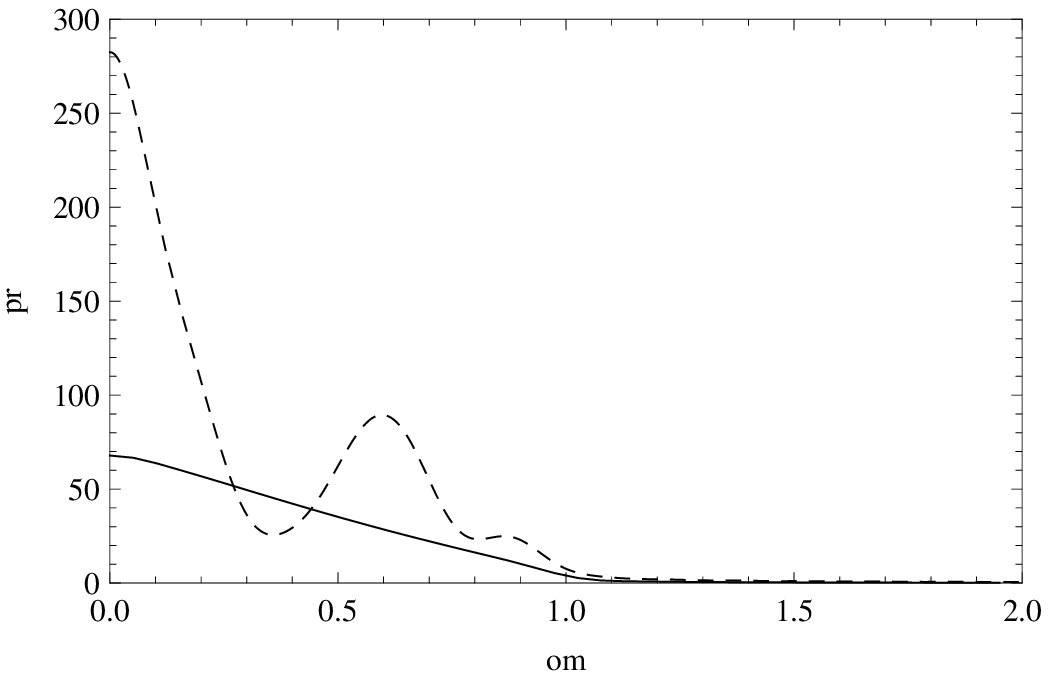}
 \caption{Upper left panel: bouncing time $T_{\om_a}^{\rm b}$ of \eqr{eq:tb} as a function of $\om/\ommax$ in unit of $\kappa^{-1}$ (points) and corresponding values of $\Gamma_a$ (boxes), when $L \kappa /\chor = 25$.
The condensate parameters are as in~\fref{fig:lorentzians}, lower plot.
Upper right panel, solid line:  transition rate associated with \eqr{rate} due to 
the radiation emitted by the sole black hole as a function of $\om/\ommax$ (solid line); corresponding quantity due to the discrete set of complex frequency modes in the black-hole--white-hole geometry.
The oscillations are due to the cosine in \eqr{eq:gamma}.
Lower panels: $P(\om, T)$ of \eqr{rate} as a function of $\om/\ommax$ for an isolated black hole (solid line) and for a black-hole--white-hole pair (dashed line), respectively at time $T=30\kappa^{-1}$ (left plot) and $T=200\kappa^{-1}$ (right plot).}
 \label{fig:fluxes}
\end{figure}

In \fref{fig:fluxes}, we represent the relevant quantities that govern the properties of the `probability' function introduced in \eqr{rate}.
On the left upper plot, we present the growth rates $\Gamma_a$ and the corresponding times $T_{\om_a}^{\rm b}$ of \eqr{eq:tb} for the 13 complex frequency modes that exist in the condensate flow of~\fref{fig:lorentzians}, lower plot.
On the right upper plot, we show the Golden Rule transition rate associated with \eqr{rate} in that black-hole--white-hole geometry (dots), and the rate one would obtain if the white hole were not present. In that case, the quantity plotted is proportional to $\om \times \bar n_\om$, where $\bar n_\om$ is the mean occupation number in the black hole geometry as computed in~\cite{mp}.
It is given by $\bar n_\om = \vert w_\om \vert^2/(1 - \vert w_\om \vert^2 )$, see $U_3$ of \eqr{eq:umatrices}, where $\vert w_\om \vert^2 $ is well approximated by
\begin{equation}
\vert w_\om \vert^2 = \ee^{- 2\pi \om/\kappa} (1 -\om/\ommax)^{1/2}.
\end{equation}
The pre-factor of $\om$ is due to the normalization of the $\chi$ modes of \eqr{chimodes}.
In a black-hole--white-hole, the quantity that corresponds to $\om \times \bar n_\om$ is $\om_a \times 2\Gamma_a T_{\om_a}^{\rm b}$. In both cases, these quantities vanish for $\om>\ommax$.

In the lower plots, we show \eqr{rate}, at two different times, and for both the black-hole--white-hole, and the isolated black hole flows. As expected, the exponential growth of the laser effect does not show up for times $T$ smaller than the inverse of the maximal $\Gamma_a$ (here $T \sim 110/\kappa$).
Moreover, the discreteness of the spectrum is not visible either at early times.
To be resolved, it requires times larger than $2\pi \Delta \om_a$, where $\Delta \om_a$ is the frequency gap between neighboring Bohr--Sommerfeld frequencies $\om_a$. It is here of the order of $\Delta \om_a \sim \ommax/10 \sim 0.02 \kappa$.

On the contrary, for times of the order of $1/{\max}(\Gamma_a)$ or greater, both the exponential growth of the laser effect and the discreteness of the spectrum show up. These allow one to distinguish the phonon flux emitted by the black-hole--white-hole pair from that emitted by the sole black hole that grows linearly in $T$ for all values of $\om$.
This linear growth can be observed by comparing the continuous lines of the right upper and lower plots, and constitutes a numerical validation of the Golden Rule!

\subsection{Spatial properties of complex frequency modes and correlation patterns}
\label{sec:modesandcorrelation}

The density--density correlation function can be calculated following the procedure described in~\sref{sec:correlations}. As outlined there, only the largest-$\Gamma_a$ mode significantly contributes to the pattern at late time.
Nevertheless, to appreciate the variety of cases, it is worth studying the modes and the corresponding correlation patterns also for lower $\Gamma$.

\begin{figure}
  \centering
   \psfrag{x}[c][c]{\small $x\kappa/\chor$}
 \psfrag{mode}[c][c]{\small $\xl$}
  \includegraphics[width=.7\textwidth]{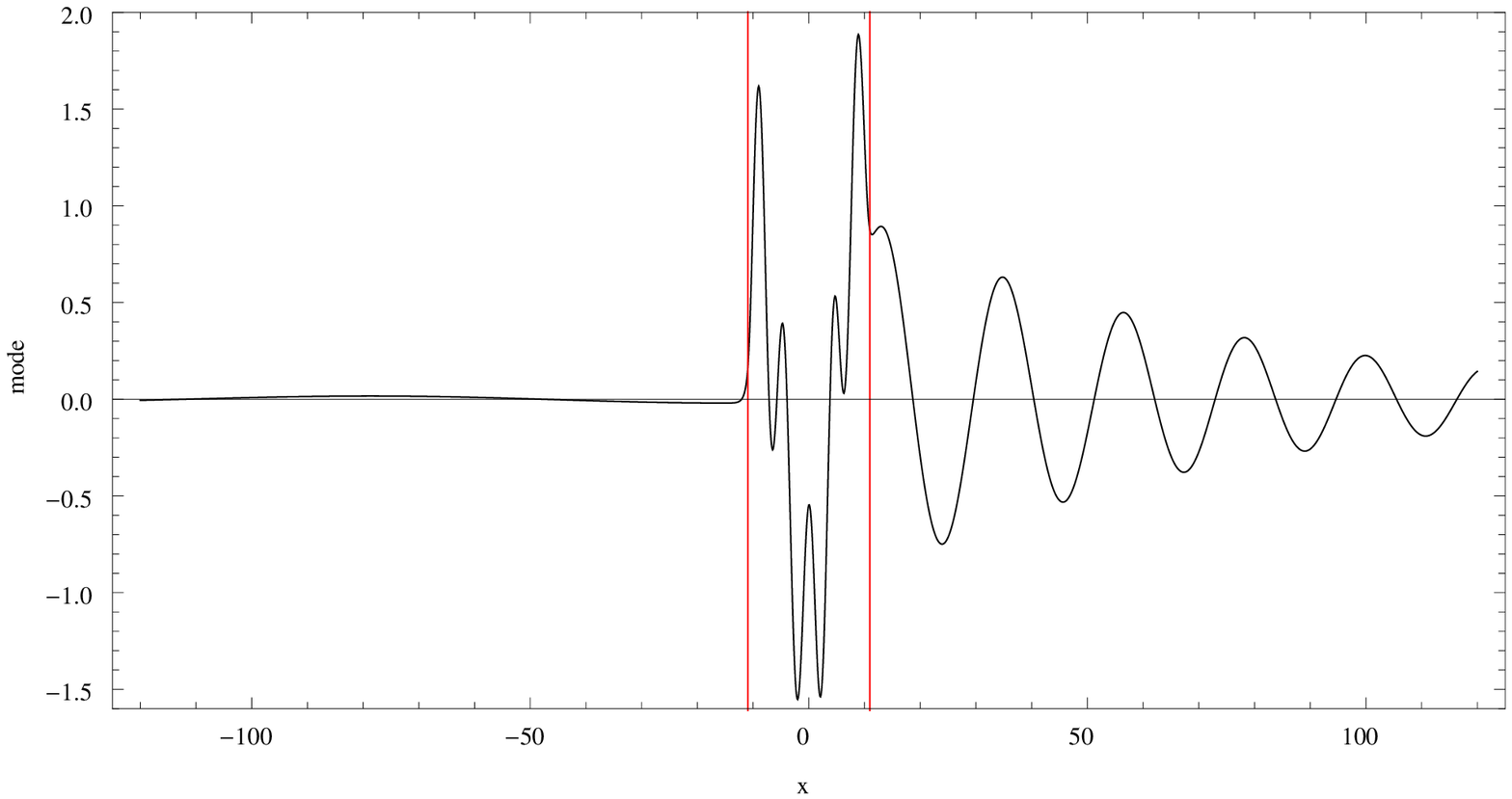}
  \includegraphics[width=.7\textwidth]{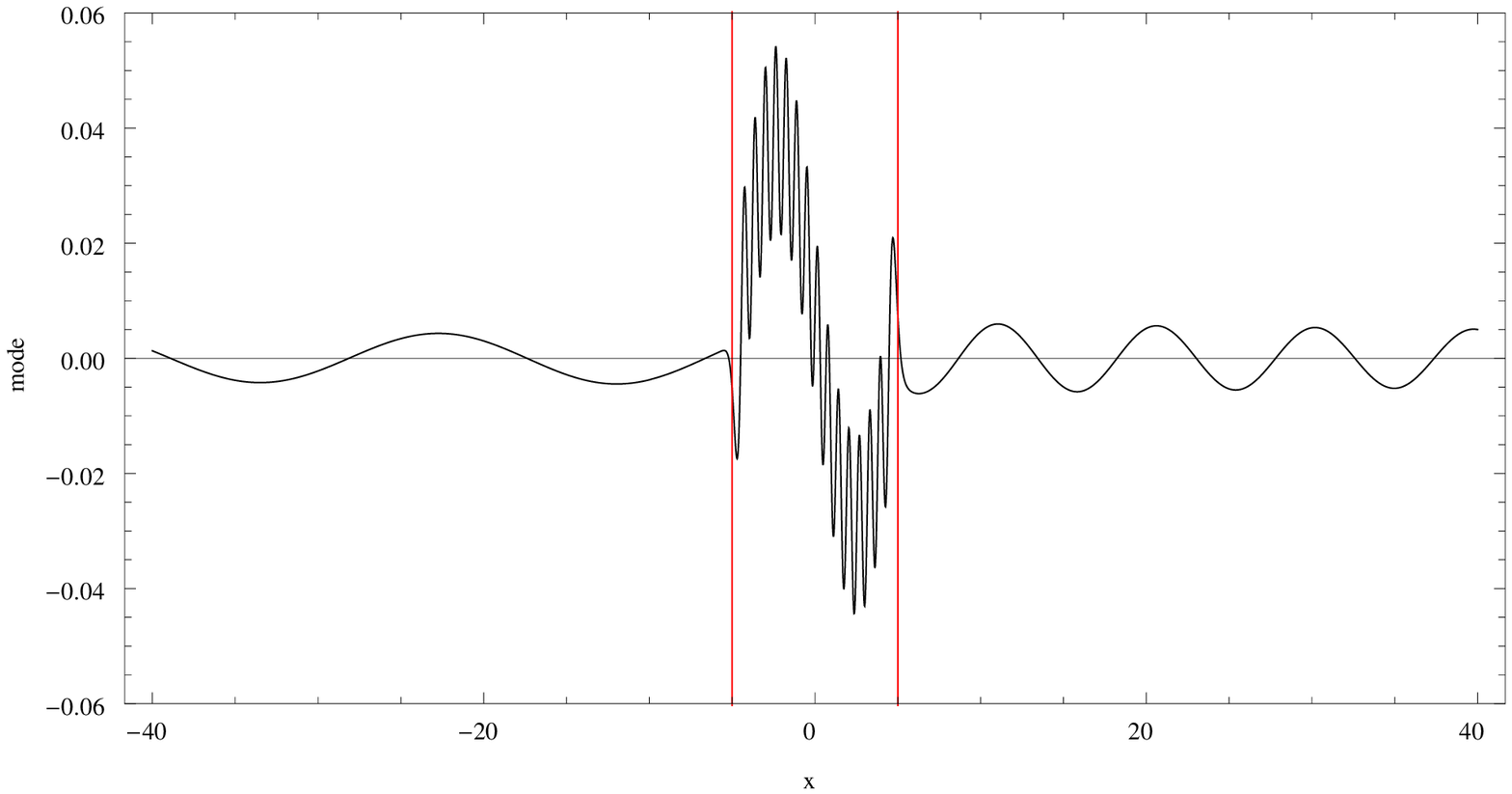}\\
  \hspace{0.01\textwidth}\includegraphics[width=.68\textwidth]{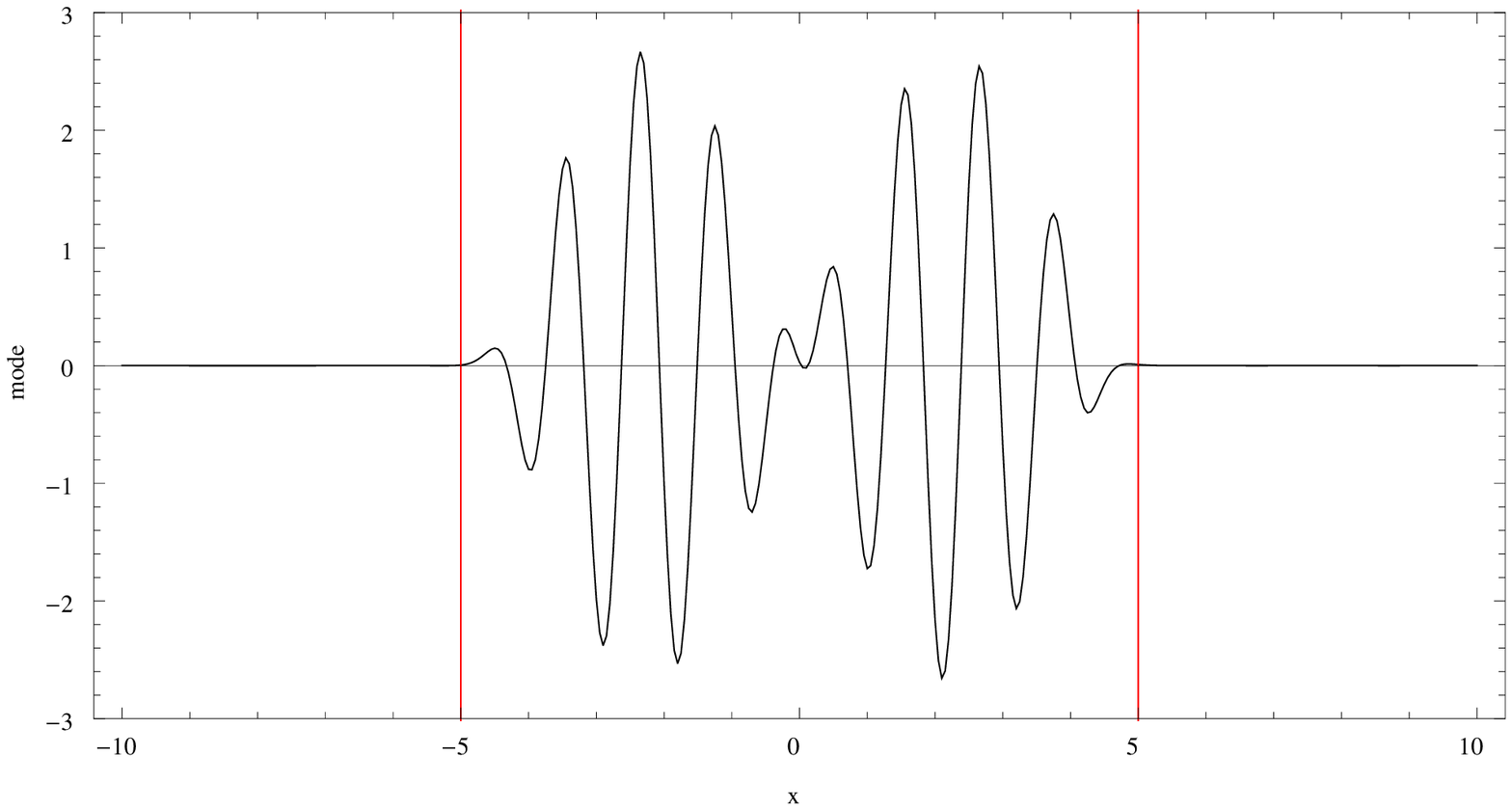}
  \caption{Real part of the growing mode $\xl$ of \eqr{eq:phiexpansion} for respectively high, moderate and low values of $\Gamma$.
Red lines: white and black horizons.
Upper panel: fourth eigenmode with $\om_4/\kappa=0.098$, $\Gamma_4/\kappa=0.0056$, with $L\kappa/\chor=10.95$ and the other parameters as those of \fref{fig:lorentzians}.
Central panel: eigenmode with $\nbs=14$, $\om_{14}/\kappa=0.46$ and $\Gamma_{14}/\kappa=0.004$.
$\ommax/\kappa=2.53$, $q=0.7$, $D=0.7$, $n=2$, $\kw=\kb=\kappa$, $\Lambda/\kappa=8$, $L\kappa/\chor=5$.
Lower panel: eigenmode in the same background with $\nbs = 2$, $\om_2/\kappa=2.48$, $\Gamma_2/\kappa=2.\times 10^{-6}$.
The importance of $\Gamma$ can be estimated from the relative mode amplitude in the right outside region versus that inside in the trapped region.
Online movies showing the evolution of $\xl$ as a function of $t$, respectively \href{http://people.sissa.it/finazzi/bec_bhlasers/movies/laser_1.gif}{laser\_1.gif} (from $t=-400/\kappa$ to $400/\kappa$), \href{http://people.sissa.it/finazzi/bec_bhlasers/movies/laser_2.gif}{laser\_2.gif} (from $t=-200/\kappa$ to $200/\kappa$), \href{http://people.sissa.it/finazzi/bec_bhlasers/movies/laser_3.gif}{laser\_3.gif} (from $t=-400/\kappa$ to $400/\kappa$)(available from~\urlnjp).
}
 \label{fig:mode}
\end{figure}

As a typical example of a mode with a large $\Gamma$, we consider the fourth BS mode in a configuration with fivecomplex eigenfrequencies. In~\fref{fig:mode} (upper panel) the real part of this mode is represented, while the supplementary movie \href{http://people.sissa.it/finazzi/bec_bhlasers/movies/laser_1.gif}{laser\_1.gif} (available from~\urlnjp) gives its evolution in time.
As a second example, we choose a very different situation with $q=0.7$ and a dispersion relation with higher $\Lambda$.
In this case, the spectrum of complex frequencies is large. In~\fref{fig:mode} two modes are plotted: one with $\nbs=14$ and a moderate value of $\Gamma$ (central panel), and one with $\nbs=2$ and a very small value of $\Gamma$ (lower panel).
When comparing the modes and their correlation patterns, some features remain the same, whereas others significantly differ.

\begin{figure}
  \centering
  \includegraphics[width=.9\textwidth]{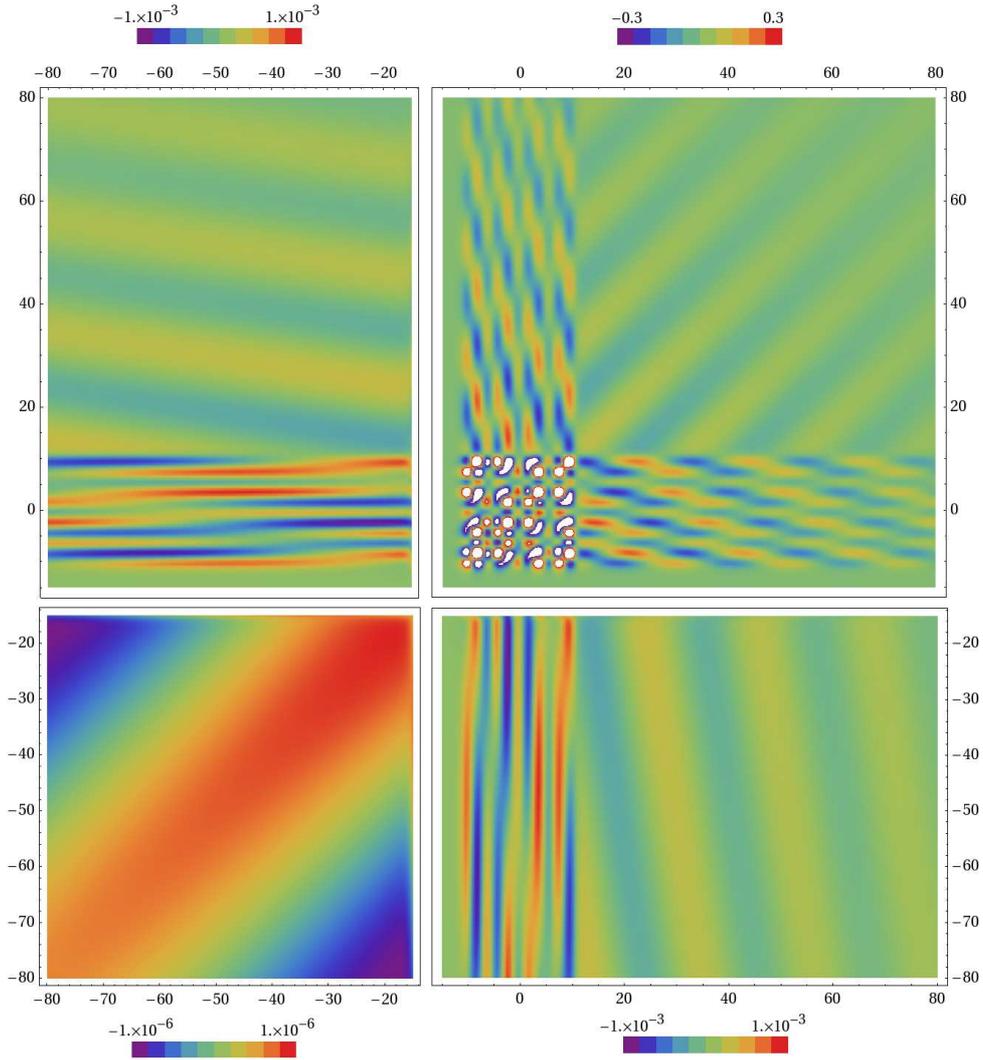}
 \caption{Density--density correlation function~\eref{eq:densitycorrelation} divided by $e^{2\Gamma_a t}$ of the mode with a high $\Gamma$ represented in the upper plot of \fref{fig:mode} with $x$ in units of $\chor/\kappa$. We have divided the expression by $e^{2\Gamma_a t}$ in order to suppress the dependence on time.
Since the correlations bewteen different regions have very different amplitudes, we were obliged to use different scales to reveal them. The central square from $-11$ till $11$ is the trapped region.
It contains the strongest correlations since the mode amplitude is highest in it.
The two (symmetrical) bands of width $\sim 22$ in the right upper plot describe the correlations between the trapped region and the external region on the right of the black hole, while the other two bands describe those with the left external region. The correlations in the lower left panel are those on the left of the white hole.
They are much weaker and of greater wave length, in agreement with the mode properties on the left side of the upper plot of \fref{fig:mode}.}
 \label{fig:correlations}
\end{figure}
\begin{figure}
  \centering
  \includegraphics[width=.9\textwidth]{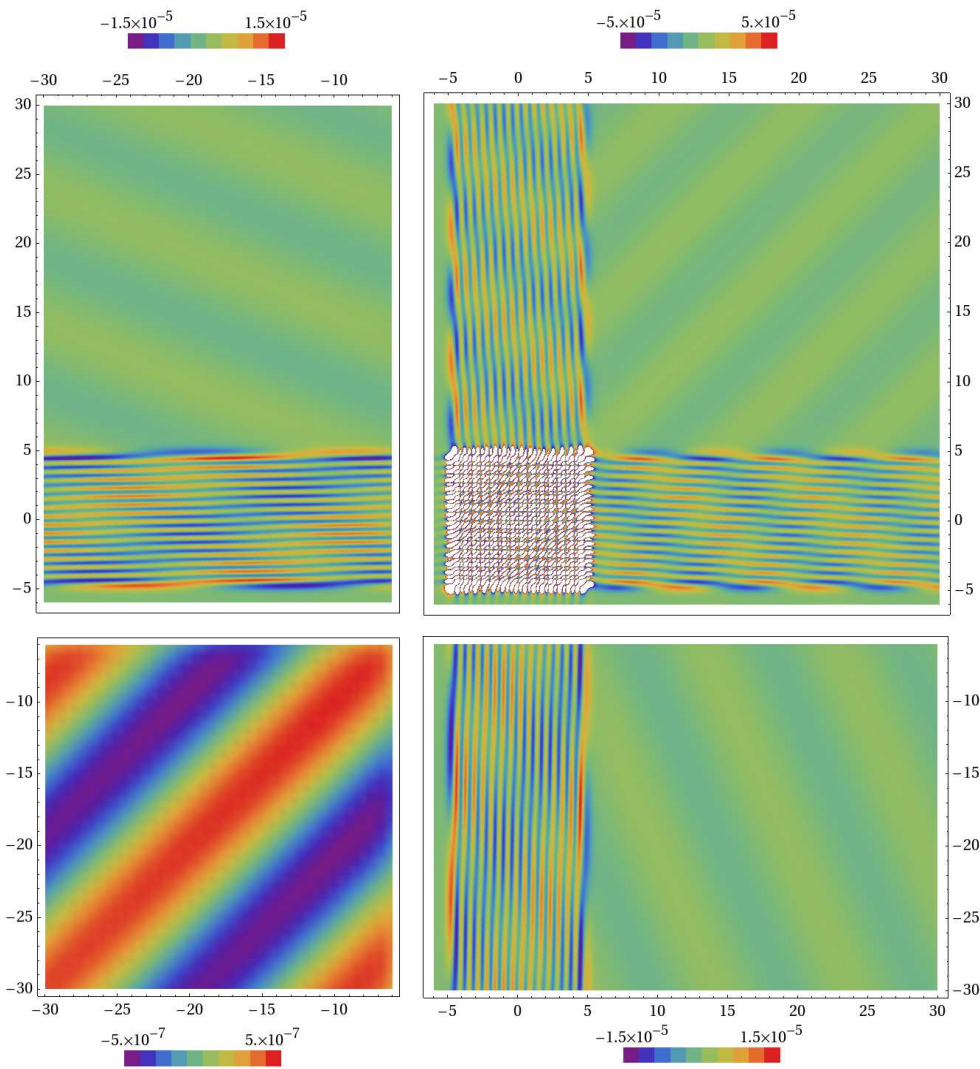}
 \caption{Density--density correlation function~\eref{eq:densitycorrelation} divided by $e^{2\Gamma_a t}$ for the mode with a moderate $\Gamma$ represented in the central plot of \fref{fig:mode}.
Similarities with the former figure are manifest, and concern both the spatial properties of the pattern
and the amplitudes of the correlations.
The main difference concerns the high-frequency modulations of the inside--inside and inside--outside correlations
which are due to the fast oscillations of the mode amplitude in the trapped region, which are visible in the central plot of \fref{fig:mode}, and which are due to fact that $\nbs$ is high: $\nbs = 14$.}
 \label{fig:correlations3}
\end{figure}
\begin{figure}
  \centering
  \includegraphics[width=\textwidth]{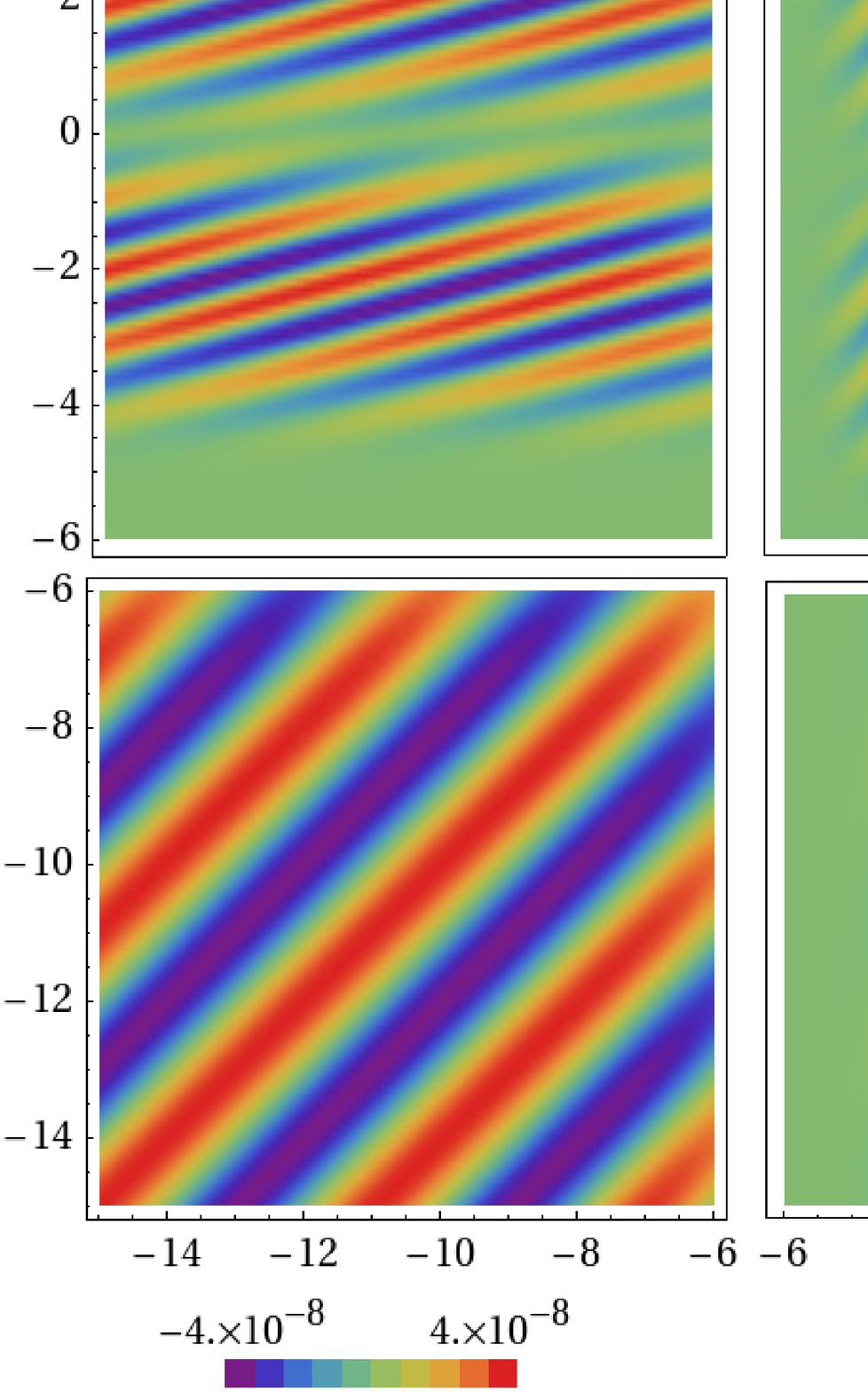}
 \caption{Density--density correlation function~\eref{eq:densitycorrelation} divided by $e^{2\Gamma_a t}$ for the mode with a low $\Gamma$ represented in the lowest panel of \fref{fig:mode}.
The bottom legend corresponds to the central region of the plot. One clearly sees the two nodes in the trapped region.
One also finds them in the inside--outside correlations, both on the left and on the right of the trapped region.
Finally, one notices that in the present case the $u-u$ correlations on the right upper square are weaker than the $v-v$ correlations on the lower left panel. This follows from the high $u$--$v$ coupling between the
trapped mode, which is a $u$-mode, and the outside $v$-mode.}
 \label{fig:correlations2}
\end{figure}

In the first example, since $\Gamma$ is high, the laser effect is strong as can be seen from the movie: in the right asymptotic region there is an exponentially growing right-going flux.
This growth in time can also be inferred from the spatial decrease of the mode amplitude on the right of the black hole, because these are linearly related, see equation~(54) of \cite{cp}.
Considering the correlation pattern of this mode in~\fref{fig:correlations}, one sees that the dominant signals come from the inside region (the central square of size $\sim 22$) where the amplitude of the mode is highest, and from the horizontal and vertical bands which describe the correlations between the escaping flux and the trapped mode.

From the mode itself and from its correlation pattern in~\fref{fig:correlations}, one can see that the $u$--$v$ mixing is very low, as expected from the choice of $q=0.5$. The left-going mode coming out of the white hole is about 2.5 orders of magnitudes smaller than the right-going one that comes out of the black hole, as can be seen from the ratio of amplitudes between the top left part and top right part in~\fref{fig:correlations}.
To understand how the modes propagate, it is appropriate to consider the top left part.
This region represents the correlations between $u$ (right-going) and $v$ (left-going) modes. The slope of the highest/lowest correlation lines gives the ratio between the velocities of the two modes. 
When the modes form a continuous set, the two velocities are the group velocities. 
However, in the present case, when considering a single mode,
the slope of these lines corresponds instead to the ratio between the phase velocities of $u$- and $v$-modes.
Similarly, the pattern in the top right part are simply due to the nodes of the right moving mode evaluated at a given time,
see equation~(61) in \cite{cp}.

We now consider the correlation pattern of~\fref{fig:correlations3}, corresponding to a mode with a moderate value of $\Gamma$ and $\nbs$ very large. The overall properties of the pattern are basically the same as those of \fref{fig:correlations}. The main differences come from the high wave-vector content of the trapped mode, which produces short distance features in the central square and in the bands. Outside these regions the patterns are very similar.

In the last pattern of~\fref{fig:correlations2}, corresponding to a mode with a very small value of $\Gamma$ and $\nbs=2$, we get a different picture. The pattern is basically concentrated in the central square since the amplitude of the mode outside is related to its amplitude inside by the square root of $\Gamma$, as can be seen from~\eref{eq:gamma}. In the present case, the $u$--$v$ mixing is so high that the pattern on the left of the white hole is higher (by a factor of about 5) than on the right of the black hole.
One also notices that in the central panel ($\nbs=2$) there is only one node. Moreover, the beat-like shape shows that this wave is the result of the superposition of two waves with very similar and relatively high wave vectors.

\subsection{The Technion experiment}
\label{sec:technion}

Recently a black-hole--white-hole flow has been experimentally realized~\cite{tech}.
Our numerical code can describe this configuration, up to some limitation.
In fact, the velocity profile (figure 3(b) of~\cite{tech}) is rather irregular while our code works accurately only for velocity profiles as in~\fref{fig:qD}, right panel, with an almost-flat internal region.
Nevertheless, we shall proceed in order to obtain at least an estimation on the number of the unstable modes and their growing rate.
As far as we know, this is the first estimate of this quantity, which is very relevant from an experimental standpoint.

The main parameters of the experimental setting are
\begin{eqnarray}
 \chor = 7.3\times10^{-4}\;{\rm m/s},\nonumber \\
 \kb = \kw/2 = \kappa = 213.7\;{\rm s}^{-1},\nonumber \\
 v(x=0) = -3.6\times10^{-3}\;{\rm m/s} = -4.9\,\chor,\nonumber \\
 c(x=0) = 3.5\times10^{-4}\;{\rm m/s} = 0.48\,\chor,\label{eq:c0}\\
 c(x\to\infty) = 8\times10^{-4}\;{\rm m/s} = 1.1\,\chor,\label{eq:cinf} \\
 2L = 10^{-5}\;{\rm m}=1.5\,\chor/\kappa,\nonumber \\
 \Lambda/\kappa=6.\nonumber 
\end{eqnarray}

\begin{figure}
 \centering
 \psfrag{om}[c][c]{\small $\omega/\ommax$}
 \psfrag{b2}[c][c]{\small $|{\cal B}_\om^{(2)}|^2$}
 \includegraphics[width=0.45\textwidth]{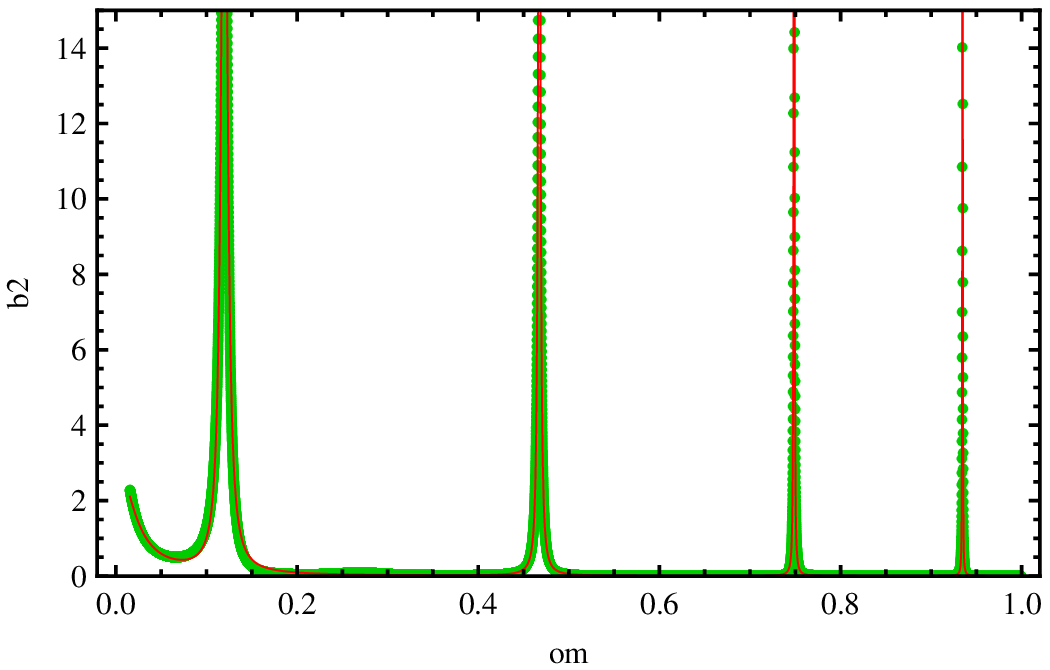}
 \includegraphics[width=0.45\textwidth]{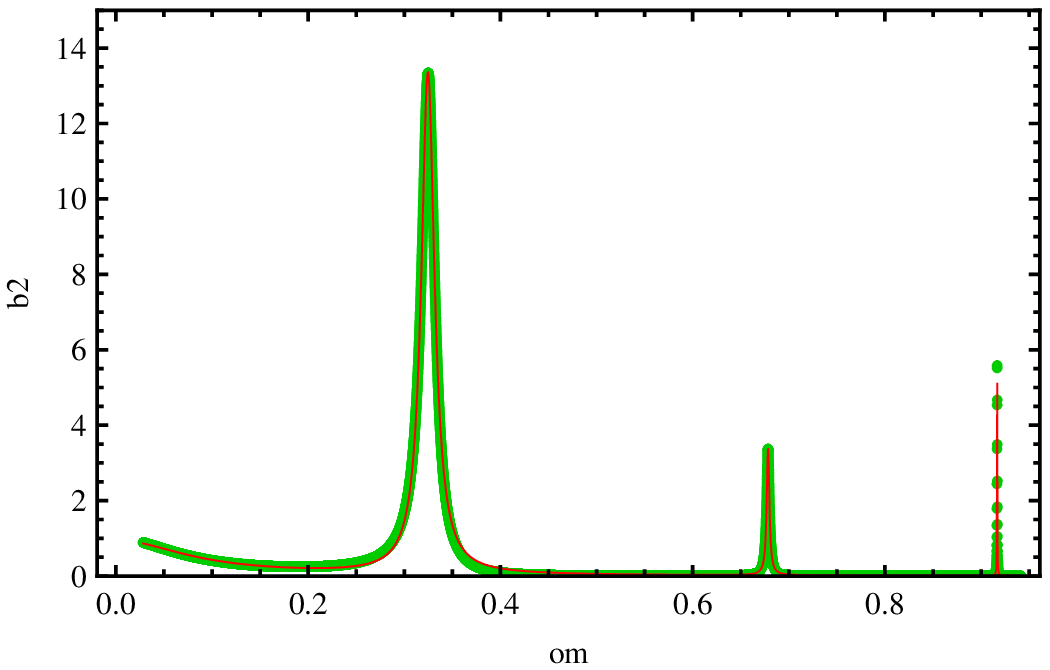}
 \caption{$|{\cal B}_\om^{(2)}|^2$ as a function of $\omega/\ommax$ for  $q=0.9$, $\ommax/\kappa\approx3.4$ (left panel) and $q=0.48$, $\ommax/\kappa\approx2.7$ (right panel).
The values of the other parameters are given in the text. Green points: numerical simulation; red lines: fitted series of Lorentzians.
$\omega_a$ and $\Gamma_a$ from the fit are reported in~\tref{tab:tech}.}
 \label{fig:tech}
\end{figure}

To reproduce the above setting we perform two simulations with fixed $D=1$.
In~\fref{fig:tech}, $|{\cal B}_\om^{(2)}|^2$ is plotted as a function of $\omega/\ommax$ for $q=0.9$ (left panel) and $q=0.48$ (right panel). These two values are obtained by using, respectively, $c(x=0)$ and $c(x\to\infty)$ equal to their experimental values in~\eref{eq:c0} and~\eref{eq:cinf}.%
\footnote{We used these two values because, in our analysis, it is not possible to fix independently $c(x=0)$ and $c(x\to\infty)$.
We expect that the actual value of the complex frequencies would lie in between the two sets we obtain.}
The small distance between the two horizon implies that few trapped modes are present in this situation.
The values of $\Gamma_a$ and $\omega_a$ of the complex eigenfrequencies are reported in both cases in~\tref{tab:tech}.
The dominant contribution comes from the eigenfrequency with the largest $\Gamma_a$.
In the two cases this corresponds to an instability time scale, respectively $\tau\approx0.06\;{\rm s}$ and $\tau\approx0.02\;{\rm s}$. Even if the two simulations are generated with very different values of $q$, the two results agree in order of magnitude. Note that the estimated instability time scale is more than twice the time scale for which the horizons are maintained in the experimental configuration ($\approx0.008\;{\rm s}$).
In order to see the laser effect, it would be needed to maintain the systems for a longer time, or to increase $\kappa$ by a factor of 10, without changing the adimensional parameters of the system such as the ratio $L\kappa/\chor$.

\Table{\label{tab:tech}$\omega_a$ and $\Gamma_a$ for the complex eigenfrequencies in the systems described in the caption of~\fref{fig:tech}.}
 \br
     \centre{3}{$q=0.9$}			&		\centre{3}{$q=0.48$}		\cr
 \mr
 $\omega_a/\ommax$ & $\omega_a/\kappa$ & $\Gamma_a/\kappa$ & $\omega_a/\ommax$ & $\omega_a/\kappa$ & $\Gamma_a/\kappa$ \cr
 \mr
 $2\times10^{-9}$ & $7\times10^{-9}$ & 0.076 & $2\times10^{-8}$ & $5\times10^{-8}$ & 0.23 \cr
 0.12 & 0.41 & 0.012 & 0.32 & 0.86 & 0.022 \cr
 0.47 & 1.58 & 0.005 & 0.68 & 1.80 & 0.004 \cr
 0.75 & 2.54 & 0.0007 & 0.92 & 2.43 & 0.0005 \cr
 0.93 & 3.17 & 0.0004 & & & \cr
 \br
 \endtab

\section{Conclusions}
\label{sec:conclusions}

In this paper, we analysed the spectrum of phonons in a BEC when the stationary flow of the condensate crosses twice the speed of sound. Our analysis is based on the Bogoliubov--de Gennes equation~\eqr{eq:hp}. Hence, even though the analogy with light propagation in a pair of black and white horizons is manifest, none of our results rely on this gravitational analogy.

In the limit where the condensate can be considered as infinite, i.e. when no periodic boundary condition is introduced, the spectrum of bound modes contains real frequency modes that are only elastically scattered, see \eqr{RT}, plus a discrete and finite set of pairs of complex frequency modes~\eqr{eq:phiexpansion}. These modes can be seen as the resonances of the cavity bordered by the two sonic horizons. They lead to dynamical instabilities because the scattering through each sonic horizon is anomalous, in that it mixes positive and negative norm modes, see~\eqr{eq:expnov}.
Then, using semi-classical techniques, we compute the real and the imaginary part of the complex frequencies. The real part $\om_a$ obeys a Bohr--Sommerfeld condition that introduces the discreteness of the spectrum, whereas the imaginary part $\Gamma_a$ is related to the norm of the scattering coefficients across the horizons, see~\eqr{eq:gamma}.
In the case of toroidal geometry, the analysis would instead be complicated by the discreteness of the wave vectors.
The instability appears only when one $\om_a$ approximately matches one of the frequencies associated with that discrete set of wave vectors. This explains the presence of the instability bands found in~\cite{Garay,Gardiner}.

In~\sref{sec:numerics}, we numerically solved the Bogoliubov--de Gennes equation for real and complex frequencies. By comparing the numerical properties with those derived using the Bohr--Sommerfeld and~\eqr{eq:expnov}, we validated the use of semi-classical methods, see figures \ref{fig:lorentzians}--\ref{fig:omega}. In~\sref{sec:growthofn} we numerically studied the growth of number of phonons emitted by the black-hole--white-hole system. In spite of the discrete character of the trapped modes, at early time, this quantity behaves very similarly to the flux that the black hole horizon would emit in the absence of the white hole horizon. Instead, for larger time both the instability and the discreteness show up. Finally, we studied the equal time correlation pattern of density--density fluctuations associated with three different complex frequency modes, respectively with high, moderate and low value of the imaginary part $\Gamma_a$.
In all cases, the instability shows up most clearly in the supersonic region where the amplitude of the trapped mode is the highest one. The correlations between the trapped mode and the emitted modes display very specific properties, see \fref{fig:correlations} and then next two ones.

Finally, we studied (within some limitations) the experimental situation realized in June 2009 in the Technion~\cite{tech}.
The results are summarized in \fref{fig:tech} and show that few unstable modes are found, and that the instability time scale is about ten times larger than the life time of the condensate.
This means that an increase by a factor of ten of the surface gravity (without changing the adimensional parameters of the system such as the ratio $L\kappa/\chor$) would lead to comparable time scales. The laser effect could then be observable.

\ack
\addcontentsline{toc}{section}{Acknowledgments}
The authors wish to thank J Macher for providing the code used in \cite{mp} and for explanations about it.
They are also grateful to A Coutant and S Liberati for stimulating discussions and useful comments.
Finally they thank J Steinhauer for providing details about the Technion experiments, and I Carusotto for interesting comments on the manuscript.

\appendix 



\section*{\hypertarget{app:motion}{Appendix A.} The Hamiltonian of density perturbations}
\addcontentsline{toc}{section}{Appendix A. The Hamiltonian of density perturbations}
\addtocounter{section}{1}

We expand $\hH$ of \eqr{eq:Hsc} in powers of $\hp$ defined in~\eref{eq:defphi} to the second order:
\begin{eqnarray}
 \fl H_0	= 
 		 \int\!\dx \,\Pn^*\left[- \hbm\pdx^2+V+\frac{g}{2}\rn\right]\Pn, \label{eq:h0}\\
 \fl \hH_1	
 	= \int\!\dx \,\Pn^*\hpd\left[- \hbm\pdx^2+V+g\rn\right]\Pn+\mbox{h.c.}, \label{eq:h1}\\
 \fl \hH_2	
 	= \int\! \dx\,\rn\left\{\hpd\left[T_\rho-\ii v\hbar\pdx-\hbm\frac{\pdx^2\Pn}{\Pn}+V+2g\rn\right]\hp + g\frac{\rn}{2}\left(\hpds+\hp^2\right)\right\},
\end{eqnarray}
where $T_\rho$ is defined in \eqr{eq:Trho}. Expanding the Heisenberg equation of motion for $\hP$,
\begin{equation}
 \ii\hbar\pdt\hP(t,x)=\comm{\hP(t,x)}{\hH},
\end{equation}
one obtains
\begin{eqnarray}
 \ii\hbar\pdt\Pn &=& \Pn\comm{\hp}{\hH_1},\\
 \ii\hbar\pdt\hp &=& \comm{\hp}{\hH_2}-\ii\hbar\hp\frac{\pdt\Pn}{\Pn}.\label{eq:ptp}
\end{eqnarray}
The former one gives the Gross--Pitaevskii equation \eqr{eq:GP}.
Defining
\begin{equation}
 \delta\hH\equiv\int\!\dx\,\rn\left[-\hbm\frac{\pdx^2\Pn}{\Pn}+V+g\rn\right]\hpd\hp,
\end{equation}
equation~\eref{eq:ptp} can be rewritten as $ \ii\hbar\pdt\hp = \comm{\hp}{\hHe}$, where
\begin{equation}
\fl \hHe\equiv\hH_2-\delta\hH=\int\! \dx\,\rn\left\{\hpd\left[T_\rho-\ii v\hbar\pdx +g\rn\right]\hp 
 + g\frac{\rn}{2}\left(\hpds+\hp^2\right)\right\}.
\end{equation}
Remarkably, it is independent of external potential $V$ and of ${\pdx^2\Pn}/{\Pn}$. When symmetrizing it with respect to $\hp$ and $\hpd$, one obtains \eqr{eq:heffsym} which is manifestly Hermitian.

\section*{\hypertarget{app:quantization}{Appendix B.} The quantization procedure}
\addcontentsline{toc}{section}{Appendix B. The quantization procedure}
\addtocounter{section}{1}

To understand how to expand the complex field operator $\hp$ in modes and creation/destruction operators, 
it is convenient to define a two-component field~\cite{ulf}
\begin{equation}
 \hW\equiv\spin{\hp}{\hpd}.
\end{equation}
Then~\eqref{eq:hp} becomes
\begin{eqnarray}
 & \ii\hbar\pdt\hW = B\hW,\label{eq:spineq}\\
 & B = (T_\rho+g\rn)\sigma_3 -\ii v \pdx +\ii g\rn\sigma_2,
\end{eqnarray}
where $\sigma_i$ are the Pauli matrices
\begin{equation}
 \sigma_1=
  \pmatrix{
  0 & 1 \cr
  1 & 0
 },
 \quad
 \sigma_2=
 \pmatrix{
  0 & -\ii\cr
  \ii & 0
 },
 \quad
 \sigma_3=
 \pmatrix{
  1 & 0\cr
  0 & -1
 }.
\end{equation}
%
%
Since the field $\hW$ is invariant under the conjugation operation defined by
\begin{equation}
 \bar{S}\equiv\sigma_1 S^\dagger,
\end{equation}
the structure of $\hW$ must be
\begin{equation}
 \hW=\sum_n (W_n \ha_n + \bar W_n \had_n),
\label{eq:hW}
\end{equation}
where $W_n$ are doublets of $\mathbb{C}$-functions, and $\sum_n$ denotes the summation over a (possibly continuous) complete sets of modes.

Using $\pdx^\tra=-\pdx$, $T_\rho^\tra=T_\rho$ and the properties of Pauli matrices, one verifies that the scalar product
\begin{equation}\label{eq:scalar}
 \scal{W_1}{W_2}\equiv\int\!\dx\,\rnx \, W_1^{*}(t,x)\sigma_3 W_2(t,x)
\end{equation}
is conserved under time evolution when $W_i$ are solution of~\eqref{eq:spineq}, since
\begin{equation}
 B^{*\tra}\sigma_3=\sigma_3 B. 
\end{equation}
%

For later convenience, we state here some properties of the scalar product that can be verified using the anticommutation relation of Pauli matrices and~\eref{eq:commphi}:
\begin{eqnarray}
 \scal{W_1}{W_2}=\scal{W_2}{W_1}^*,\\
 \scal{\bar W_1}{\bar W_2}=-\scal{W_2}{W_1}=-\scal{W_1}{W_2}^*,\\
 \hW\hW^\tra-(\hW\hW^\tra)^\tra=\frac{1}{\rnx}\delta(x-x')\ii\sigma_2,\\
 \comm{\scal{W_1}{\hW}}{\scal{W_2}{\hW}}=-\scal{W_1}{\bar W_2}=\scal{W_2}{\bar W_1}.\label{eq:commfor}
\end{eqnarray}

Assuming that the $W_n$ have a non-zero norm, we define an orthonormal basis:
\begin{eqnarray}
 \scal{W_n}{W_m} &=& -\scal{\bar W_n}{\bar W_m}=\delta_{nm},\label{eq:wnorm}\\
 \scal{W_n}{\bar W_m} &=& 0,
\end{eqnarray}
where the Kronecker $\delta$ is replaced by a $\delta$-distribution in the case of a continuous set of modes.
Using \eqr{eq:hW} and the above identities, one obtains
\begin{equation}
 \comm{\ha_n}{\had_{m}}=\comm{\scal{W_n}{\hW}}{-\scal{\bar W_m}{\hW}}=\scal{W_n}{W_m}=\delta_{nm},
\end{equation}
which shows that $\ha_n$ and $\had_n$ are in fact destruction and creation operators.

When the condensate is stationary, one can work with eigenmodes $W_\lambda^\alpha$ of frequency $\lambda$, where the index $\alpha$ describes the set of modes with the same frequency.
By definition, one has
\begin{equation}
 B W_\lambda^\alpha=\hbar\lambda W_\lambda^\alpha.
\end{equation}
The conservation of the scalar product gives
\begin{equation}\label{eq:nocomplex}
 0=\pdt\scal{W_\lambda^\alpha}{W_{\lambda'}^{\alpha'}}=-\ii(\lambda^*-\lambda')\scal{W_\lambda^\alpha}{W_{\lambda'}^{\alpha'}}.
\end{equation}
From this it is clear that only real frequency modes can be normalized as in \eref{eq:wnorm}.
However, in the presence of dynamical instabilities, complex frequency eigenmodes are present.
We shall assume that these modes form a discrete and finite set of pairs of modes
with conjugated frequencies that we call $\{\lambda_a=\omega_a+\ii\Gamma_a, a=1, 2, ... N\}$.
In place of~\eref{eq:wnorm}, we choose the following pseudo-normalization for each pair,
\begin{equation}
 \scal{W_{\lambda_a}}{W_{\lambda_{a'}^*}}=\ii\delta_{aa'},
\end{equation}
and the other scalar product must vanish because of~\eref{eq:nocomplex}.
In fact, the Hermiticity of $\hH$ implies that if $\lambda_a$ is an eigenfrequency, then $\lambda_a^*$ is an eigenfrequency too~\cite{ulf}.
We call $V_{\lambda_a}$ and $Z_{\lambda_a}$ the doublets corresponding to $\lambda_a$ and $\lambda_a^*$, respectively.

Summarizing, the field $\hW$ can be expanded as
\begin{equation}\label{eq:fieldexpansion}
 \hW=\int\!\dom\sum_{\alpha}\!\left[\wom\aom+\womb\aomd\right]+\sum_a\!\left[\vl\bl+\zl\cl+\vlb\bld+\zlb\cld\right],
\end{equation}
where all $\lambda_a$ have a positive imaginary part $\Gamma_a$. The modes satisfy the following normalization rules,
\begin{eqnarray}
 \scal{\wom}{\womp}&=&-\scal{\womb}{\wombp}=\delta(\omega-\omega')\delta_{\alpha\alpha'},\label{eq:normWom}\\
 \scal{\vl}{\zlp}&=&\scal{\vlb}{\zlbp}=-\scal{\zlp}{\vl}=-\scal{\zlbp}{\vlb}=\ii\delta_{aa'},\label{eq:normVZ}
\end{eqnarray}
and all the other scalar product vanish.
Using
\begin{eqnarray}
 \bl&=&\ii\scal{\zl}{\hW},\\
 \bld&=&\ii\scal{\zlb}{\hW},\\
 \cl&=&-\ii\scal{\vl}{\hW},\\
 \cld&=&-\ii\scal{\vlb}{\hW}
\end{eqnarray}
and~\eref{eq:commfor}, one obtains
\begin{eqnarray}
 \comm{\bl}{\cldp}&=&\comm{\ii\scal{\zl}{\hW}}{-\ii\scal{\vlbp}{\hW}}=-\scal{\zl}{\vlp}=\ii\delta_{a a'},
\end{eqnarray}
which is the commutation relation of complex unstable oscillators~\cite{cp}.
Decomposing the doublets as
\begin{equation}
 \fl \wom=\phm{\om t}\spin{\pom}{\vpom},\quad V=\phm{\la t}\spin{\xl}{\el},\quad Z=\phm{\las t}\spin{\psl}{\zel},
\end{equation}
the field $\hp$ can be expanded as~\eqref{eq:phiexpansion}, and equations~\eref{eq:normWom} and~\eref{eq:normVZ} give \eref{eq:normphi} and \eref{eq:normpsi}.

Using~\eqref{eq:phiexpansion}, $\hHe$ is a sum of three terms, containing respectively only real frequency modes, only complex frequency modes and mixing real and complex frequencies:
\begin{equation}
\hHe =\hHr+\hHc+\hHm.
\end{equation}
Using the commutation relations and the scalar products of~\sref{sec:quantization}, we obtain
\begin{eqnarray}
 \fl\hHr &=& \frac{\hbar}{2} \int\!\dom\dom'\sum_{\alpha\alpha'}\int\!\dx\,\rnx
 \left[
  (\omp+\om)\php{\om t}\poms\phm{\omp t}\pomp\aomd\aomp
  \right. \nonumber \\ \fl&& \left.
 +(\om-\omp)\php{\om t}\poms\php{\omp t}\vpomsp\aomd\aomdp
  +(\om'-\om)\phm{\om t}\vpom\phm{\omp t}\pomp\aom\aomp
  \right. \nonumber \\ \fl&& \left.
 -(\omp+\om)\phm{\om t}\vpom\php{\omp t}\vpomsp\aomd\aomp
 \right]
 \nonumber \\
 \fl&=&\frac{\hbar}{2}\int\!\dom\dom'\sum_{\alpha\alpha'} (\om+\omp) \scal{\wom}{\womp} \aomd\aomp
 \nonumber \\
 \fl&&-\frac{\hbar}{2}\int\!\dom\dom'\sum_{\alpha\alpha'} (\om+\omp) \delta(\om-\omp)\delta_{\alpha\alpha'}
 \int\!\dx\,\rnx \phm{(\omp-\om)t}(\vpoms\vpomp)
  \nonumber \\
\fl &&+\frac{\hbar}{2}\int\!\dom\dom'\sum_{\alpha\alpha'} \om \scal{\wom}{\wombp} \aomd\aomdp
 +\frac{\hbar}{2}\int\!\dom\dom'\sum_{\alpha\alpha'} \om \scal{\womb}{\womp} \aom\aomp
 \nonumber \\
 \fl&=&\int\!\dom\sum_\alpha\hbar\om\left[\aomd\aom-\int\!\dx\,\rnx|\vpom|^2  \right].
\end{eqnarray}
Using the same techniques, it is easy to show that $\hHm=0$ and
\begin{equation}
 \hHc=\ii\hbar\sum_a \left\{\las\!\left[\bld\cl\!+\!\!\int\!\dx\rnx\zel\els\right]- \mbox{h.c.} 
\right\}\!.
\end{equation}
Putting everything together, one obtains \eqr{eq:hamiltonianabc}.

Note that $\hb$'s and $\hc$'s are not destruction operators [see~\eqref{eq:combcd}].
Nevertheless, they can be combined to give couples of creation/destruction operators
\begin{equation}
 \dlpl\equiv\ppl\frac{\bl+\ii\cl}{\sqrt{2}},\qquad \dlmi\equiv\pmi\frac{\bld+\ii\cld}{\sqrt{2}},
\end{equation}
which in fact satisfy the commutation relations
\begin{equation}
 \comm{\dlpl}{\dlpldp}=\comm{\dlmi}{\dlmidp}=\delta_{aa'},
\end{equation}
and all the other commutators vanish. Equation~\eqref{eq:fieldexpansion} becomes
\begin{equation}
 \fl \hW=\int\!\dom\sum_{\alpha}\left[\wom\aom+\womb\aomd\right]+\sum_a\left[\wlpl\dlpl+\wlmi\dlmi+\wlplb\dlpld+\wlmib\dlmid\right],
\end{equation}
where
\begin{equation}
 \wlpl\equiv\pplm\frac{\vl-\ii\zl}{\sqrt{2}},\qquad\wlmi\equiv\pmim\frac{\vlb-\ii\zlb}{\sqrt{2}},
\end{equation}
whose normalization is
\begin{equation}
\scal{\wlpl}{\wlplp}=\!\scal{\wlmi}{\wlmip}=\!-\scal{\wlplb}{\wlplbp}=\!-\scal{\wlmib}{\wlmibp}=\!\delta_{aa'},
\end{equation}
and the other scalar products vanish. Notice that $W_{a\pm}$ are no longer frequency eigenmodes.
Decomposing $\wlpl$ and $\wlmi$ as
\begin{equation}
 \wlpl=\spin{\plpl}{\vplpl},\quad \wlmi=\spin{\plmi}{\vplmi},
\end{equation}
the field expansion~\eqref{eq:phiexpansion} becomes~\eqref{eq:phiadd}, and the Hamiltonian~\eqref{eq:hamiltonianabc} becomes~\eqref{eq:hamiltonianadd} when the (arbitrary) phases $\theta_\pm$ are chosen to be $\theta_\pm=0$.

\section*{References}
\addcontentsline{toc}{section}{References}

\end{document}